\documentclass[onecolumn,nobibnotes,nofootinbib]{revtex4}
\usepackage{subfigure}
\usepackage{epsfig}
\usepackage{multirow}
\usepackage{color}
\usepackage{hyperref}
\usepackage{array}
\usepackage{float}
\usepackage{url}
\usepackage{changepage}
\usepackage{orcidlink}
\usepackage{amsmath,amssymb,bm}
\usepackage{appendix}
\newcommand{\be}{\begin{eqnarray}}
\newcommand{\ee}{\end{eqnarray}}

\makeatletter
\newcommand{\thickhline}{%
  \noalign {\ifnum 0=`}\fi \hrule height 1.2pt
  \futurelet \reserved@a \@xhline
}
\newcolumntype{"}{@{\hskip\tabcolsep\vrule width 1.2pt\hskip\tabcolsep}}
\makeatother
\begin{document}
\title{Testing the double-logarithm asymptotic gluon density in ultraperipheral heavy ion collisions at the Large Hadron Collider}

\author{D. A. Fagundes\orcidlink{0000-0002-3504-3671}} 
\email[E-mail me at: ]{daniel.fagundes@ufsc.br}
\affiliation{Department of Exact Sciences and Education, CEE. Federal University of Santa Catarina (UFSC) - Blumenau Campus, 89065-300, Blumenau, SC, Brazil}

\author{M.V.T. Machado\orcidlink{0000-0003-2821-8266}}
\email[E-mail me at: ]{magno.machado@ufrgs.br}
\affiliation{High Energy Physics Phenomenology Group, GFPAE. Institute of Physics, Federal University of Rio Grande do Sul (UFRGS) Postal Code 15051, CEP 91501-970, Porto Alegre, RS, Brazil}

\begin{abstract}
In this   paper, we analyze the application of an analytical gluon distribution based on double-asymptotic scaling to the photoproduction of vector mesons in coherent $pp$, $pA$, and $AA$ collisions at LHC energies, using the color dipole formalism. Predictions for the rapidity distribution are presented for $\rho^0$, $J/ \psi$, $\psi (2S)$, and $\Upsilon (1S)$ mesons photoproduction. An analysis of the uncertainties associated with different implementations of the dipole--proton amplitude is performed. The vector meson 
photoproduction accompanied by electromagnetic dissociation is also analyzed.
\end{abstract}

\keywords{exclusive vector meson photoproduction; relativistic  ultraperipheral collision; nuclear electromagnetic  dissociation; QCD color dipole formalism}
%\PACS{PACS numbers come here}
%
%%%
%
\maketitle
%

%%% Text
%
\section{Introduction}
\label{sec:intro}

The exclusive production of vector mesons plays {a crucial %an important
} %MDPI: colloquial, not recommended for scientific papers. Please consider the the replacing. here and elsewhere below. 
%R.: WE AGREE WITH THE SUGGESTION.
 role both experimentally and theoretically, enabling accurate studies on applications of perturbative quantum chromodynamics (QCD) and parton saturation physics. Light vector mesons, on one hand, have no 
hard {(of transverse momentum above 1 GeV)} %MDPI: added to clarify to the readers, please confirm the meaning is retained.
%R.: WE AGREE WITH THE SUGGESTION. MEANING RETAINED.
perturbative scale associated with the process in the photoproduction limit, 
{with the incoming photon's virtuality 
 %MDPI: to be defined as met first. Please confirm the intended meaning has been retained. No undefined abbreviations, variables, quantities, indexes etc. as met first in each of the main parts of a paper such as abstract, mainbody text, figures, tables even known. Please consider. Here and elsewhere below.
 %R.: WE AGREE WITH THE SUGGESTION. MEANING RETAINED.
$Q^2=0$,}  
thus probing the non-perturbative regime of QCD. In the electroproduction case, {where %the virtuality %MDPI: alredy defined. No repetition of the definitions. Please consider. 
%R.: WE AGREE WITH THE SUGGESTION.
$Q^2$}  %of the incoming photon 
is sufficiently large, a perturbative approach is justified. On the other hand, quarkonium production involves an intrinsic hard scale characterized by the heavy quark masses, even at $Q^2=0$, and weak coupling methods can be successfully applied. 
 It is {%well 
 known} %MDPI: No "well/wide/etc-known/used/etc" recommended for scientific papers. Please consider. Here and elsewhere below.
 %R.: WE AGREE WITH THE SUGGESTION. 
hl{transverse momentum below 1 GeV} %MDPI: clarifies to the readers. Please confirm the meaning is retatined. 
%R.: WE AGREE WITH THE SUGGESTION. MEANING RETAINED.
that the transition between the perturbative QCD treatment and the soft regime can be handled by the parton saturation 
framework~\cite{Gribov:1983ivg,Mueller:1985wy,Mueller:2001fv,McLerran:1993ni,McLerran:1993ka,McLerran:1994vd,Ayala:1995kg,Ayala:1995hx,Iancu:2003xm,Gelis:2010nm,Morreale:2021pnn} 
within the QCD color dipole formalism~\cite{Nikolaev:1990ja,Nikolaev:1991et,Mueller:1993rr,Mueller:1994jq}. There, the saturation scale, $Q_s$, sets the limit on the 
parton phase-space density that can be reached in the hadron’s wavefunction. Hence, it can be viewed as the typical underlying scale in 
 {low %MDPI: No "small", "big", "large", "fast", "slow", "near", "low"  etc.  with no comparison, estimate recommended for scientific papers. Please consider. Here and elsewhere below.
 %R.: WE AGREE WITH THE SUGGESTION. 
 enough} %MDPI: or provide the limit in number. 
 Bjorken-$x$ dynamics for light meson photoproduction {(with $x$ a fraction of the hadron's longitudinal momentum carried by a parton)}  %MDPI: Please confirm 
%R.: WE AGREE WITH THE SUGGESTION.
the    
and also plays a key role in quarkonium production (see recent applications in Refs.~\cite{Henkels:2022bne,Cisek:2022yjj,Mantysaari:2022sux,Mantysaari:2022kdm,Mantysaari:2022bsp,Mantysaari:2021ryb,Kopeliovich:2021dgx,Kopeliovich:2022jwe,Cepila:2023pvh,Matousek:2022enl,Kumar:2022aly,Anand:2018zle,Xie:2022sjm,Henkels:2023plt,Mantysaari:2023xcu,Mantysaari:2023prg,Azevedo:2023fee,Goncalves:2023sts,Boroun:2023vuu,Boroun:2023klw} and references therein). The 
photon--proton {($\gamma p$)} or photon--nucleus {($\gamma A$)} 
scattering in such an approach is characterized by the QCD color dipole amplitude, which represents the interaction of 
the 
Fock states of the incoming virtual photon, i.e., color singlet dipoles at leading order, with the target at rest. The 
 {quark--anti-quark 
($q\bar{q}$)}  
fluctuations and the transition from the dipole to meson are described by the photon and vector meson wavefunctions, respectively.

Given the shutdown of the Hadron-Electron Ring Accelerator (HERA) experiment, low-$x$ studies of exclusive processes 
 are now being actively pursued in relativistic heavy-ion {collisions %(RHICs) %MDPI: RHIC denotes Relativistic Heavy Ion Collider, not collisions. Please consider.
 %R.: WE AGREE WITH THE SUGGESTION. MEANING RETAINED.
 and} at the Large Hadron Collider (LHC). In 
ultraperipheral heavy-ion collisions (UPCs) and 
coherent proton--proton ($pp$) reactions, the production process can be theoretically factorized into the high photon flux from the hadron or nucleus projectile and the photon--target production cross-section 
 {
 %(see Refs. 
\cite{Bertulani:1987tz,Baur:2001jj,Bertulani:2005ru,Baltz:2007kq,Klein:2020fmr}%)
}. %MDPI: fits better, no parenthses necessary, clear enough.  
%R.: WE AGREE WITH THE SUGGESTION. MEANING RETAINED.
Investigations into $\gamma p$ or $\gamma A$ interactions at these colliders provide valuable information about the underlying QCD dynamics. A key piece of information that can be extracted from the available meson production data is the gluon distribution function, 
$xg(x,\mu^2)${, where $\mu$ is the characteristic hard scale}. %MDPI: defined. Please confirm the intended meaning retaimned. 
%R.: WE AGREE WITH THE SUGGESTION. MEANING RETAINED.

Recently, in Ref.~\cite{Fagundes:2022bzw}, the gluon parton distribution function (PDF) was evaluated by us based on double-asymptotic scaling (DAS) for gluon density~\cite{Caola:2008xr,Ball:2016spl} within the color dipole approach. The model has the advantage of replacing the numerical LO DGLAP evolution at large virtualities with analytical calculations from pQCD in the DAS approximation. The evolution presents similar results compared to other studies~\cite{Rezaeian:2013tka,Mantysaari:2018nng} using numerical DGLAP evolution. 
{Reference} %MDPI: the full-wording as starting a sentence. 
%R.: WE AGREE WITH THE SUGGESTION. 
~\cite{Fagundes:2022bzw} contributes to {a considerably} %an 
{easier} %MDPI: No "simple/ly", "easy/ily", "perfect/ly", "beautiful/ly" etc. recommended for scientific papers. Please consider replacing or removal. Here and elsewhere below.
%R.: WE AGREE WITH THE SUGGESTION. MEANING RETAINED.
modeling of the dipole cross-section and requires {only %very %MDPI: No "very", "extreme/ly", "huge/ly", "vast/ly", "too", "much", "great/ly" etc. recommended for scientific papers. Please consider replacing. Here and elsewhere below.
%R.: WE AGREE WITH THE SUGGESTION. MEANING RETAINED.
} 
 {a little} computational power to predict inclusive and exclusive deep inelastic scattering (DIS) observables. Regarding vector meson production, the calculations of 
$J/\psi$ and $\rho$ {mesons} 
photoproduction cross-sections describe both the HERA data and the measurements of vector meson photoproduction in %proton--proton ( %MDPI: already defined.
%R.: WE AGREE WITH THE SUGGESTION. 
$pp$ %) 
collisions at the LHC. Moreover, as shown in Ref.~\cite{Fagundes:2022bzw}, the extracted analytical gluon density is quite suitable for the high-precision HERA data 
and for studying small-$x$ physics at future colliders such as the Electron Ion Collider (EIC) and the Large Hadron Electron \mbox{Collider (LHeC).}

The main focus {of the current paper} %MDPI: fits better, makes it clearer. Please consider. 
%R.: WE AGREE WITH THE SUGGESTION. 
 %here 
is to investigate the robustness of the phenomenological model for the color dipole amplitude based on DAS approximation as applied to UPCs and $pp$ collisions at the LHC energies. We analyze the theoretical uncertainties on the predictions of exclusive production of vector mesons in 
 %proton--proton, proton--lead, and lead--lead 
 {$pp$, $p$Pb, PbPb} %MDPI: fits better.Please consider.   
 %R.: WE AGREE WITH THE SUGGESTION. 
 collisions at the LHC, providing predictions for the rapidity distribution, $y$. This paper is organized as follows. In Section~\ref{sec:2}, we present the color dipole scattering amplitude that incorporates QCD evolution through the DAS approximation. The parameters of the model were determined from small-$x$ high-precision DIS 
 {electron--proton 
($ep$) }  
data; the model correctly describes the structure functions and exclusive vector meson production. In Section~\ref{sec:3}, we apply the formalism in order to compute the 
production of vector mesons in UPCs, focusing on the LHC energy regime. In the last Section, {the %main %MDPI: fits better, otherwise why not to discuss al results is unclear, suspicious. 
%R.: WE AGREE WITH THE SUGGESTION. 
results obtained} are discussed.

\section{Theoretical framework and phenomenological models}
\label{sec:2}

Let us start by defining the ansatz for the gluon density based on the DAS approximation investigated in Ref.~\cite{Fagundes:2022bzw}. Given the sufficiently soft behavior at the input scale, $Q_0$, one obtains the double-asymptotic scaling for the gluon density as 
 {follows}%MDPI: 1. suggested to replace here the $\gamma$ with, e.g., $\kappa$ to not mismatch with the photon nomenclature, $\rho$ with, e.g., %$\lambda$ to not mismatch with ro=meson nomnclature. 
 %R.: WE PREFER THE ORIGINAL NOTATION. LET IT AS IT IS.
 %2. The parenthese and the dividion symbol are scaled. 
 %R.: WE AGREE WITH THE SUGGESTION. 
 %3. In eq(1) the squared bracketd=s replaced with the parenthses. Please consider. 
 %R.: WE AGREE WITH THE SUGGESTION. 
~\cite{Caola:2008xr,Ball:2016spl}:
\vspace{-3pt}\begin{eqnarray}
xg_{\mathrm{DAS}}(x,\mu^2) & = & \frac{A_g}{\sqrt{4\pi \gamma \sigma}}\exp \left(%[ 
2\gamma \sigma - \delta \frac{\sigma}{\rho} \right)%]
, \label{eq:model1}\\
 \gamma & = & \sqrt{12/\beta_0}, \quad \delta = \left(11+\frac{2N_f}{27}\right)\Big/\beta_0 ,
\end{eqnarray}
{where} %$N_c$ denotes the number of colors and %MDPI: N_c is not used in eqs(1), (2), so moved to Eq(8) where it comes first. 
%R.: WE AGREE WITH THE SUGGESTION. MEANING RETAINED.
{$A_g$} %MDPI: moved here from below eq5 as met it first here. Please consider.
%R.: WE AGREE WITH THE SUGGESTION. MEANING RETAINED.
denotes the overall normalization of the distribution and  
 $\beta_0 = 11 -\frac{2}{3}N_f$ ({with $N_f$ %is 
 the} number of active flavors) denotes the first term of the QCD beta-function series. The geometric mean $\sigma$ 
and the ratio $\rho$ define the double-scaling variables:
\begin{eqnarray}
 \sigma & \equiv & \sqrt{\ln\left( \frac{x_0}{x} \right) \ln \left( \frac{t}{t_0} \right) }, \,\,\,
\rho \equiv \sqrt{\ln\left( \frac{x_0}{x} \right)\Big/\ln \left( \frac{t}{t_0} \right)},\\
 t/t_0 & \equiv & \ln (\mu^2/\Lambda_{\mathrm{QCD}}^2)/\ln (Q_0^2/\Lambda_{\mathrm{QCD}}^2),  
\end{eqnarray}
where the parameters $x_0\simeq 10^{-1}$ and $Q_0^2\simeq 1$ GeV$^2$ define the boundaries of the asymptotic domain, and 
 %$A_g$ denotes the overall normalization of the distribution 
 {$\Lambda_{\mathrm{QCD}}$ represents the QCD scale parameter} %MDPI: clarifies, please confirm the meaning is retained.
 %R.: WE AGREE WITH THE SUGGESTION. MEANING RETAINED.
\cite{Ball:1994du}. Accordingly, the $xg$ exhibits asymptotic scaling in both $\sigma$ and $\rho$ in the double limit 
of {large $\sigma$ ($\sigma \rightarrow \infty)$} %MDPI: highly helpful to provide the value in the parenthese, e.g., (sigma > ...) or say "large enough", "relatively large" if suitable. 
 at fixed but reasonably large $\rho$ and {large   
$\rho$ ($\rho \rightarrow \infty$)} %MDPI: highly helpful to provide the value in the parenthese, e.g., (rho > X) or say "large enough", "relatively large" if suitable.
at fixed $\sigma$. The original DAS ansatz provides a {exceptionally %very 
 good} %MDPI: No "good", "well", "nice" etc.  quality estimate recommended for scientific papers. Please consider replacing. Here and elsewhere below.
 %R.: WE AGREE WITH THE SUGGESTION. MEANING RETAINED.
 description of DIS data, as demonstrated by the last three rows of \mbox{Table \ref{tab:Model1_bin1_hqs}}, which follows from our previous {study %work
}~\cite{Fagundes:2022bzw}.

In Ref.~\cite{Fagundes:2022bzw}, the form in Equation (\ref{eq:model1}) was modified, enabling the smooth transition from 
{large ($x\rightarrow 1$)} %MDPI: would be helpful to provide the value in the parentheses, e.g., (above X)
to {lower ($x\rightarrow 0$)} %MDPI: would be helpful to provide the value in the parentheses, e.g., (below X) 
values of $x$ in $xg$. Namely, in the modified DAS model, the following form was {considered:} %MDPI: the squared brackets moved to the parenthses. Please consider.  
%R.: WE AGREE WITH THE SUGGESTION. 
\begin{eqnarray}
\label{eq:model2.2}
xg_{\mathrm{FM}}(x,\mu^2) &=& xg(x,\mu^{2}_{0})\exp \left(%[ 
2\gamma \sigma^{\prime} - \delta \frac{\sigma}{\rho^{\prime}} \right)%] 
 , \\
 {\rm with}\, \,\, xg(x,\mu^{2}_{0})&=&A_{g}x^{-\lambda_g}(1-x)^{C_g},\nonumber
\end{eqnarray}
where $xg(x,\mu^{2}_{0})${, with $ \mu_{0}$ the initial evolution scale,}  %MDPI: 1. please define. 
%R.: Ok, defined mu0
%2. there is NO mu_0 in the expression xg(x,mu_0), please check and clarify. 
%R.: $xg(x,\mu^{2}_{0})$ is the gluon dendity ansatz at the initial scale. IT IS CORRECT.
%stands for 
represents 
the soft gluon %\textit 
{ansatz}  %, MDPI: no the italics necessary. Please consider. The italics is reserved for subsection, book titles, journal volumes, etc. or the wording to be stressed. 
%R.: YES, AGREE.
(with %\linebreak 
$C_{g}=6.0${, $\lambda_g$ is the low-$x$ effective power)}, %MDPI: please define or provide the number.
%R.: YES, IT IS NOW DEFINED.
and the rescaled geometric mean and ratio are now given by $\sigma^{\prime}= \sigma \ell (x)$ and $\rho^{\prime}= \rho \ell (x)$, respectively.
The function $\ell (x) = \sqrt{\mathcal{N}(1-x)^{5}}$, in which $\mathcal{N}$ represents a new free parameter of the model, controls the normalization of $xg(x,\mu^2)$ with the evolution of $\mu^2$. Such a factor corrects both the overall gluon normalization and the transition to larger $x$ in the asymptotic term $e^{2\gamma \sigma}$ within the \mbox{original expression. }

\begin{table}[H]
\centering
\caption{{Fit} %MDPI: We have adjusted the format of the table and removed vertical lines and inner horizontal lines, please consider. 
%R.: YES, AGREE.
 parameters of the DAS model in Equation~(\ref{eq:model1}) for {$Q^{2}\in%: %MDPI: \in fits better. Please consider. 
%R.: YES, AGREE.
(1.5,50)$} GeV$^2$ and $x\leqslant 0.001$, including charm and bottom quarks. All fit parameters are provided at a $70\%$ confidence level, with 
{the masses} \mbox{$m_{lq}=0.03$} GeV {of light quarks}%MDPI: defines the parameters shown. Please confirm the intended meaning is retained.  
%R.: YES, AGREE.
, $m_{c}=1.3$ GeV {of $c$ quark}, and $m_{b}=4.2$ GeV {of  $b$ quark}, while 
{%$C_g=4.0$ being fixed throughout and 
$\mu_{0}$ is defined in Equation} {\eqref{eq:model2.2}
and $\sigma_0$ is defined in Equation \eqref{BGK-satDIP}
}%MDPI: please comnfirm the parameter, otherwise very unclear: ther is NO C in Eq(1). Please check.
.}

\vspace{.3cm}
\resizebox{.9\textwidth}{!}{
\begin{tabular}{c|c|c|c|c|c|c}
\thickhline
\textbf{Model} & $\boldsymbol{\sigma}_\textbf{0}$ \textbf{[mb]} & $\textbf{\emph{A}}_{\textbf{\emph{g}}}$ & $\textbf{\emph{x}}_{\textbf{0}}$ & $\boldsymbol{\mu}_{\textbf{0}}^{\textbf{2}}$ \textbf{[GeV$^{2}$]} & $\boldsymbol{\chi}^\textbf{2}$\textbf{/dof} & $\textbf{\emph{p}}$\textbf{-Value}\\
\thickhline
\multirow{4}{*}{BGK} &  $ (172.4\pm 2.9)\times 10^{5} $  & $  1.624\pm 0.0041 $  & 1.0 (fixed) & 1.1 (fixed) &  $1485.02/288=5.16  $ &  0  \\
\cline{2-7}
 &  $ 295 \pm 29$  & $ 2.834 \pm 0.017 $  & 0.1 (fixed) & 1.1 (fixed) &  $306.848/288= 1.07 $ &  0.213  \\
 \cline{2-7}
& $194 \pm 17$  & $3.047 \pm 0.074$  & $0.0805\pm 0.0078 $  &  1.1 (fixed) &  $303.262/287=1.06$ &  0.244  \\
\cline{2-7}
& $114 \pm 11$  & $3.80 \pm 0.32$  & $0.0496 \pm 0.0095$  &  $1.29 \pm 0.13 $ &  $290.576/286=1.02$ & 0.414\\   
\thickhline
\end{tabular}
}
\label{tab:Model1_bin1_hqs}
\end{table}

Given the DAS approximation, the QCD color dipole amplitude can be obtained. In the target rest frame, the DIS $ep$ process, $\gamma^*p\rightarrow X$, 
{where $\gamma^*$ denotes the virtual photon and $X$ denotes all produced hadrons,} is viewed as the 
interaction of a color singlet $q\bar{q}$ pair with fixed transverse size $\vec{r}$ scattering off the target through gluon exchanges (gluon ladder diagrams). For 
instance, in the impact-parameter saturation model {(IPSAT%, hereafter %MDPI: not necessary. Please consider.
%R.: YES, AGREE.
)} \cite{Kowalski:2003hm,Mantysaari:2018nng,Mantysaari:2018zdd}, which describes the multiple interactions of a QCD dipole probe with a dense target, the 
 {$b$}-dependent dipole amplitude is given {as} %by the following: 
\be
\frac{d\sigma_{q\bar{q}}}{d^2\vec{b}} = 2\,\left[1-
\exp\left(-\frac{\pi^2}{2\,N_c}r^2\alpha_S(\mu^2)xg(x,\mu^2)T(b)\right)
\right],       
\label{IP-satDIP}
\ee
where { $T(b)$ represents the thickness function of a proton,   
$N_c$ denotes the number of colors, %and 
 $\alpha_S$ represents the strong interaction coupling, 
} %MDPI: please confirm.  
%R.: YES, WE CONFIRM.
the scale is set as $\mu^2=C/r^2+\mu_0^2$, and $x$ denotes the modified Bjorken 
 {variable,} $x=x_{\mathrm{Bj}}\left(1+
 {4m_q^2}/{Q^2}\right)$, 
%\frac{4m_q^2}{Q^2}\right)$,  %MDPI: the ratio linearized as visible better in the text. Please consider. 
%R.: YES, AGREE.
with 
{$x_{\mathrm{Bj}}$ being a fraction of proton carried by the struck quark and }
$m_q$ being the effective quark mass~\cite{Golec-Biernat:1998zce,GolecBiernat:1999qd}. Such a modification consistently describes the transition from 
{high enough $Q^2$} %MDPI: please add or say "quite high", "high enough", "relatively high" if appropriate. 
toward the limit $Q^2\rightarrow 0$. Thus, the corresponding dipole cross-section is obtained as\linebreak   $\hat \sigma(x, \vec{r}) =\int d^2\vec{b}\, d\sigma_{q\bar{q}}/d^2\vec{b}$. 

%In Equation (\ref{IP-satDIP}), $T(b)$ represents the thickness function of a proton. 
Generally, a Gaussian model is assumed for $T(b)$, given the exponential fall-off in the $|t|$-dependence of quarkonia production {, 
with $t$ denoting the momentum transfer}.  %MDPI: please add the definition. 
%R.: DEFINITION ADDED.
 In such a case, the normalized-to-unity thickness, $T_G(b)$, is given {as} %by the following: 
\be
\label{TBPROTON}
T_G(b) = \frac{1}{2\pi B_{G}} \exp \left(-\frac{b^2}{2B_{G}}\right),
%\quad \int d^2\vec{b}\,T_G(b) =1,
\ee
where the parameter $B_G\simeq 4$ GeV$^2$ is related to the average squared transverse radius of the nucleon, $\langle b^2 \rangle = 2B_G$~\cite{Kowalski:2003hm,Mantysaari:2018nng,Mantysaari:2018zdd}. 

On the other hand, in the BGK approach~\cite{Bartels:2002cj,Golec-Biernat:2006koa,Golec-Biernat:2017lfv}, the color dipole cross-section 
 {reads} %MDPI: f its better. Please consider.
 %takes the following form: 
\be
\hat \sigma(x, \vec{r})= \sigma_0\,\left[1-
\exp\left(-\frac{\pi^2r^2\alpha_S(\mu^2)xg(x,\mu^2)/N_c}{\sigma_0}\right)
\right],       
\label{BGK-satDIP}
\ee
where $\sigma_0$ is now a new parameter of this model. The unintegrated gluon distribution (UGD) ${\mathcal{F}}(x,k)$% 
---{the distribution probability of a gluon 
 %having 
 with 
a transverse momentum $k$ and longitudinal momentum fraction $x$ 
to be found in the nucleons}---%MDPI: please define in wording, otherwise very unclear what it represents. %R.: DEFINITION ADDED.
can be obtained from the dipole cross-section \eqref{BGK-satDIP} %above
{as}~\cite{Luszczak:2022fkf,Boroun:2023ldq}  
%follows: 
\begin{eqnarray}
\mathcal{F}(x,k) = k^4\frac{\sigma_0}{\alpha_S(\mu_0^2)}\,\frac{N_c}{8\pi^2}  \int_0^{\infty}rdr \,J_0(kr)\left[1-\frac{\alpha_S(\mu_0^2)}{\alpha_S(r)}\,\frac{\hat 
\sigma(x, r)}{\sigma_0} \right] , %.
\end{eqnarray}
{where %$k$ denotes the gluon transverse momentum
} %MDPI: please provide the definition.
%R.: DEFINITION ADDED.{and}  
$J_0(\cdot)$ %MDPI: for functions, but is always useful to used the dotted parenthses argument to stress this is a function and not a variable; e.g. gamma use as a function and as a variable below. Please consider.
%R.: YES, WE AGREE.
represents the zeroth-order Bessel function of the first kind. %MDPI:moved here from the below eq.(17) as first met here.  please confirm.
%R.: YES, WE CONFIRM.

For the DIS-inclusive process, the cross-section for the interaction of a virtual photon with virtuality $Q^2$ with a given polarization off a proton target is expressed 
{as}~\cite{Kovchegov:2012mbw}:
\be
\label{sigtot}
\sigma_{T,L}^{\gamma^*p}(x,Q^2)= \sum_f
\int d^2\vec{r}\int dz \left|\psi_{T,L}^f(Q,r,z)\right|^2 \hat\sigma (x,\vec{r}), 
\ee
where $\psi_{T,L}^f(Q,r,z)$ denotes the corresponding photon wave function in the mixed representation for a photon with transverse, $T$, or longitudinal, $L$, 
polarization, respectively. The quark (antiquark) carries a longitudinal momentum fraction $z$ ($1-z$) of the incoming photon. The summation above refers to the quark 
flavor, $f$. The squared photon wave functions, summed over quark helicities for a given photon polarization and quark flavor, $f$, are expressed 
by~\cite{Kovchegov:2012mbw}: 
\begin{eqnarray}
 \left|\psi_{T}^f(Q,r,z)\right|^2 & = & \frac{2N_c}{\pi}\alpha_{\mathrm{\rm em}}e_f^2\left\{\left[z^2+(1-z)^2\right]\epsilon^2 K_1^2(\epsilon r) + m_f^2 K_0^2(\epsilon 
r)\right\}, \\
 \left|\psi_{L}^f(Q,r,z)\right|^2 & =& \frac{8N_c}{\pi}\alpha_{\mathrm{\rm em}}e_f^2 Q^2 z^2(1-z)^2 K_0^2(\epsilon r), %.
\end{eqnarray}
where {$\alpha_{\mathrm{\rm em}}$ %MDPI: the subscript moved to the regular (Roman/normal) font as represents the abbreviatio, not the indexes, neither a product of two variables. Please confirm. Here and elsewhere below.
is the electromagnetic coupling, $e_f$  and $m_f$ denote the $f$-flavor quark's charge  
and mass, respectively, } %MDPI: please provide/confirm the definition the definitions.  
{$K_{\nu}(\cdot)$} denotes the modified Bessel functions of the second kind of {the} order { %, 
 $\nu=0$ and $%,
 1$, } %MDPI: the commas displaced, "and" added, fit better, otherwise mismatching. Please consider. 
% R.: YES, WE AGREE. 
and $\epsilon = \sqrt{z(1-z)Q^2+m_f^2}$.

The exclusive processes can also be described within the very
 same QCD color dipole picture. For example, the differential cross-section for exclusive vector meson {($V$)} %MDPI: clarifies. Please consider.
% R.: YES, WE AGREE. 
production, $\gamma^* p \rightarrow Vp$, is given 
by~\cite{Kowalski:2006hc}:
\begin{eqnarray}
 \frac{d\sigma^{\gamma p\rightarrow Vp}}{d t} = \frac{1}{16\pi}\left|\mathcal{A}^{\gamma p\rightarrow Vp}\right|^2\;(1+\beta^2)\,R_g^2,
 \label{exclusive-meson}
\end{eqnarray}
where {$\mathcal{A}^{\gamma p\rightarrow Vp}$ represents the color dipole scattering amplitude defined below in this Section, %MDPI: please confirm. 
% R.: YES, WE AGREE. 
and } 
the real-to-imaginary ratio of the scattering amplitude, $\beta$,  
and the skewness factor, $R_g^2$, are computed as follows: 
\begin{eqnarray}
 \beta & = & \tan\left(\frac{\pi\lambda_{\mathrm{eff}}}{2}\right), \quad\text{with}\quad \lambda_{\mathrm{eff}} \equiv \frac{\partial\ln\left(\mathcal{A}_{T}^{\gamma p\rightarrow Vp}\right)}{\partial\ln(1/x)},\\
 R_g(\lambda_{\mathrm{eff}}) & =& \frac{2^{2\lambda_{\mathrm{eff}}+3}}{\sqrt{\pi}}\frac{\Gamma(\lambda_{\mathrm{eff}}+5/2)}{\Gamma(\lambda_{\mathrm{eff}}+4)},
%.
 \label{eq:RG}
\end{eqnarray}
 {where $\Gamma(\cdot)$ is the gamma function.}

{The} %MDPI: Please confirm whether the paragraph below the formula to be indented.
% R.: YES, THE PARAGRAPH HAS TO BE IDENTED.
 factor $R_g$ incorporates the off-diagonal effect, coming from the fact that the gluons attached to the $q\bar{q}$ can carry different light-front fractions $x,x^{\prime}$ of the nucleon.
The factor given in Equation~(\ref{eq:RG}) was obtained at the NLO level, in the limit that $x^{\prime}\ll x\ll 1$ and at small $t$, assuming that the diagonal gluon 
density of the proton follows a power-law form~\cite{Shuvaev:1999ce} (see the discussion in Ref.~\cite{Harland-Lang:2013xba}). This phenomenological approach has been considered %by most authors 
 {extensively} 
in the literature. %In any case, %MDPI: colloquial, not recommended for scientific papers. Please consider the replacing or removal. 
% R.: YES, WE AGREE. 
 {Anyway,} 
for {relatively} small 
dipoles (dominant for heavy{-}meson production), the dipole cross-section or amplitude is proportional to the diagonal 
gluon distribution, which {to %will
} %MDPI: No the future tenses, aiming, promising for the current paper, findings, statements etc. Please consider. here and elsewhere below. 
% R.: YES, WE AGREE. 
 be corrected by the Shuvaev formula in the off-diagonal case. In the past, the skewness factor has been extracted from the data by comparing the amplitudes for 
deeply virtual Compton scattering (DVCS) and the DIS amplitude (see, {e.g.,} %MDPI: "e.g." fits better for listing numbers, refs, etc.Please consider.
 %for instance, 
 % R.: YES, WE AGREE. 
Refs. \cite{Favart:2005sc,Favart:2007zz,Schoeffel:2008xw}). The parametrization in Equation~(\ref{eq:RG}) is %quite 
satisfactory {enough} %MDPI:  or just "satisfactory". Fits better, otherwise looks like excellent, which cann not be believable. Please consider.
% R.: YES, WE AGREE. 
 at describing it.

The elastic scattering amplitude for the process $\gamma p\rightarrow Vp$ is a function of $x$ and of the momentum transfer $\vec{\Delta}$ (with $|t|=\vec{\Delta}^2$), written as a Fourier transform of the photon and vector meson wavefunctions convoluted with the 
color dipole scattering amplitude
{transverse ($T$) or longitudinal ($L$) polarization contributions}~\cite{Kowalski:2006hc}:
\begin{eqnarray} 
 \mathcal{A}^{\gamma p\rightarrow Vp}_{T,L}& = & \mathrm{i}\,\int\!d^2\vec{r}\int_0^1\!\frac{d{z}}{4\pi}\int\!d^2\vec{b}\;(\Psi_{V}^{*}\Psi_{\gamma})_{T,L}\;\mathrm{e}^{-\mathrm{i}[\vec{b}-(1-z)\vec{r}]\cdot\vec{\Delta}}\;\frac{d\sigma_{q\bar q}}{d^2\vec{b}}, \\
 & = & \mathrm{i}\,\pi\int_0^\infty\!r\,d{r}\,\int_0^1d{z}\int_0^\infty\!b\,d{b}\,(\Psi_V^*\Psi_{\gamma})_{T}\;J_0(b\Delta)\;J_0\left([1-z]r\Delta\right)\;\frac{d\sigma_{q\bar{q}}}{d^2\vec{b}}
.%, 
\end{eqnarray}
%where $J_{0}(x)$ denotes the zeroth-order Bessel function of the first kind. %MDPI: already defined above in Eq(9).. 

The overlap function, $\Psi^{*}_{V}\Psi_{\gamma}$, between the photon and the vector{-}meson wave functions is {%well 
commonly  %MDPI: or "generally", "in general", or simply remove the charecterizng wording. Please consider the replacing.  
% R.: YES, WE AGREE. 
known}. Here, we are interested in vector{-}meson 
photoproduction, and the corresponding overlap involves only the transverse polarization contribution. In this case, the wave function at $Q^2=0$ 
%simplified 
{simplifies to} % the following:
\vspace{-5pt}

\begin{eqnarray}
 (\Psi_V^*\Psi_{\gamma})_{T} = \hat{e}_f \sqrt{4\pi\alpha_{\rm em}}\, \frac{N_c}{\pi z(1-z)} \,
 \left\{m_f^2 K_0(\epsilon r)\phi_T(r,z) - \left[z^2+(1-z)^2\right]\epsilon K_1(\epsilon r) \partial_r \phi_T(r,z)\right\},
\end{eqnarray}

where $\hat{e}_f$ denotes the effective charge{, $\phi_T(r,z)$ is the scalar part of the meson wave function in the case of transverse polarization as defined 
just next, and $\partial_a\equiv \partial / \partial_a$.}

In this study, the %\textit
{boosted Gaussian} %MDPI: please confirm the italics font is not necessary.
% R.: ITALIC NOT NECESSARY.
wave function~\cite{Kowalski:2006hc} is considered, where the scalar part of {the  
transversly or longitudinally polarized} %MDPI: please confirm; otherwise the subscripts T and L are undefined, unclear.
% R.: YES, WE AGREE. 
meson wave function $ \phi_{T,L}$ is expressed as follows. For the ground state vector meson (1S) and its first excited state (2S), one 
 \mbox{{has}~\cite{Nemchik:1994fp,Nemchik:1996cw,Cox:2009ag,Armesto:2014sma}: }
 \vspace{2pt}
\begin{eqnarray}
\label{eq:phi1s}
 \phi_{T,L}^{(1_{\rm S})}(r,z) & = & \mathcal{N}_{T,L} z(1-z)
 \exp\left(-\frac{m_f^2 \mathcal{R}^2}{8z(1-z)} - \frac{2z(1-z)r^2}{\mathcal{R}^2} + \frac{m_f^2\mathcal{R}^2}{2}\right),\\
 \phi_{T,L}^{(2{\rm S})}(r,z) & = & \mathcal{N}_{T,L}^{\prime} z(1-z)
 \exp\left(-\frac{m_f^2 \mathcal{R}^{\prime 2}}{8z(1-z)} - \frac{2z(1-z)r^2}{\mathcal{R}^{\prime 2}} + \frac{m_f^2\mathcal{R}^{\prime 2}}{2}\right)\nonumber \\
 & \times & \left[ 1+ \alpha_V \left(2+\frac{m_f^2 \mathcal{R}^{\prime 2}}{4z(1-z)} - \frac{4z(1-z)r^2}{\mathcal{R}^{\prime 2}} - m_f^2\mathcal{R}^{\prime 2}\right) \right],
 \label{eq:phi2s}
\end{eqnarray}
where the parameter $\alpha_V$ controls the position of the node of the radial wave function of the excited state.
The corresponding parameters $\mathcal{N}_{T,L}$ ($\mathcal{N}_{T,L}^{\prime}$) and $\mathcal{R}$ ($\mathcal{R}^{\prime}$) are properly obtained from both wave function normalization and the constraint from the electronic decay width, $\Gamma_{V\to e^+e^-}$. The obtained values using the quark masses of $m_f=0.03$ GeV for the $\rho$, \mbox{$m_f=1.3$ GeV} for $J/\psi$, $\psi$(2S), and $m_f=4.2$ GeV for $\varUpsilon$ are 
 {given} %presented 
in Table \ref{tab:BG_params} {along with the%. The 
~predicted} and measured decay widths of {those %these 
mesons.} %MDPI: rewritten, fits better. Please cosndider.
% R.: YES, WE AGREE. 
% are also presented.

%We are aware that %MDPI: not necessary, is a known fact. Please consider.. 
% R.: YES, WE AGREE. 
 %numerical 
 {Numerical} 
predictions for exclusive vector meson photoproduction and electroproduction depend on the choice of the vector{-}meson wave function. 
One of {the authors (M.V.T.M.)}  %MDPI: the initials can be removed as clear from teh refs, up to the author.
% R.: YES, WE AGREE. 
 %us 
has {been already involved in the analysis of } 
 %already analyzed  %MDPI: rewritten, fits better in scientific papers. Please consider.
 % R.: YES, WE AGREE. 
this 
type of theoretical uncertainty in \mbox{Refs. \cite{SampaiodosSantos:2014puz,SampaiodosSantos:2014qtt,Goncalves:2017wgg}.} 
 %Basically, 
 {Actually,} %MDPI: fits better.Please confirm the intended meaning is retained. 
 % R.: YES, WE AGREE. 
the change in the meson wavefunction parametrization, given the same model of dipole cross-section, modifies the overall normalization. 
The effect {becomes of %is more important 
 more importance} 
for lighter mesons, where the normalization variation %is 
 {becomes} 
 quite~pronounced.

\begin{table}[t]
    \centering
    \caption{Parameters of the boosted Gaussian wave function for the vector mesons analyzed. {See text for details.}}
    \begin{tabular}{c|c|c|c|c|c|c|c|c|c}
    \thickhline
         \text{\textbf{Meson}} & $\textbf{\emph{M}}_{\textbf{\emph{V}}}$ \textbf{[GeV]}& $\textbf{\emph{m}}_{\textbf{\emph{f}}}$\textbf{[GeV]} & $\boldsymbol{\mathcal{N}}_{\textbf{\emph{T}}}$ & $\boldsymbol{\mathcal{N}}_{\textbf{\emph{L}}}$ & $\boldsymbol{\mathcal{R}}$ \textbf{[GeV$^{\boldsymbol{-}\textbf{1}}$]} & $\boldsymbol{\alpha}_{\textbf{2\emph{S}}}$ & $\boldsymbol{\Gamma}_{\textbf{\emph{V}}\boldsymbol{\rightarrow} \textbf{\emph{e}}^{\boldsymbol{+}}\textbf{\emph{e}}^{\boldsymbol{-}}}^{\text{\textbf{exp}}}$ \textbf{[keV]} & $\boldsymbol{\Gamma}_{\textbf{\emph{V}}\boldsymbol{\rightarrow} \textbf{\emph{e}}^{\boldsymbol{+}}\textbf{\emph{e}}^{\boldsymbol{-}}}^{\text{\textbf{calc}}}$ \textbf{[keV]} & \textbf{Ref.} \\
         \hline
         $\rho $ &0.7753 & 0.030 & 0.9942  & 0.8928 & 3.6388 & $-$ & $ 7.04\pm 0.06$ & 7.04 & \cite{Fagundes:2022bzw}\\  
         \hline
         $J/\psi $ & 3.097 & 1.3 & 0.5974 & 0.5940 & 1.5181 & $-$ &  $5.53 \pm 0.11 $ & 5.53 & \cite{Fagundes:2022bzw}\\         
         \hline
         $\psi(\text{2S}) $ & 3.686 & 1.3 & 0.70 & 0.69 & 1.93 & -0.61 &  $ 2.33 \pm 0.08 $ & 2.35 & \cite{Armesto:2014sma} \\         
         \hline
         $\varUpsilon $  & 9.460 & 4.2 & 0.6866  & 0.6696 & 0.3354 & $-$ & $ 1.29 \pm 0.07$ & 1.29 & This study\\     
      \thickhline
    \end{tabular}

    \label{tab:BG_params}
\end{table}
\noindent

 The light photoproduction is dominated by the non-perturbative (soft) sector, and the confinement effects are significant. This feature can be embedded into the photon wave 
function by replacing its original form {with} %MDPI: We removed the squared brackets in f_s. Please confirm f_s is not an integer. 
% R.: YES, WE AGREE. f_s is not an integer.
%the following 
\cite{Frankfurt:1997zk,Forshaw:1999uf,Goncalves:2020cir} %: 
\begin{eqnarray}
 \psi_{T,L}^f(Q,r,z)\rightarrow \sqrt{f_s(r)}\,\psi_{T,L}^f(Q,r,z), \quad f_{s}(r) = %\left[
 \frac{1 + B \exp\left( -\omega^{2} (r - R)^{2} \right)}
 {1 + B \exp\left( -\omega^{2} R^{2} \right)}
 %\right]
,
 \label{factor-fs}
\end{eqnarray}
where the parameters $B$, $\omega$, and $R$ are determined by fitting the total photoproduction cross-section, $\sigma (\gamma p\rightarrow X)$. The shifted
Gaussian {function %, 
$f_s(r)$ in Equation} {\eqref{factor-fs} %, shown above,  %MDPI: no "above", "below", "mentioned", "next" etc. alone recommended but numbers of eqs, sects, etc. to better orient t he readers. Please consider.
% R.: YES, WE AGREE. 
controls} the width and height of the soft contribution enhancement for the photon wavefunction. Moreover, $f_s\rightarrow 1$ for 
{relatively} small dipoles, and {then} the hard contribution is 
unchanged. %Here,  
{In what follows, %MDPI: fits better. Please confirm the intended meaning is retained. 
% R.: YES, WE AGREE. 
we %will 
 consider} the optimal set of parameters for the %IP-SAT 
 {IPSAT}  
model in the case of $\rho$ photoproduction{:} %MDPI: clarifies. Please consider.
% R.: YES, WE AGREE. 
$B = -0.75$, $\omega = 0.25$ GeV, and $R = 6.8$ GeV$^{-1}$ 
(fixed)~\cite{Goncalves:2020cir}. 

Let us move now to the nuclear color dipole amplitude. In this {paper%work
, we consider} %MDPI: fits better, makes it clearer addressing the current study. Please confirm the intended meaning is retained. 
% R.: YES, WE AGREE. 
the Glauber--Gribov approach %is considered
~\cite{glauber1959lectures,Gribov:1968jf,Gribov:1968gs}.  
Therefore, the amplitude describing the interaction of the color dipole with a {large enough} $(A>40)$ %MDPI: would be useful to provide the number as possible: (A>X).
nucleus at %the 
{a fixed impact parameter, $b$,} %MDPI: the letter was used for the proton thickness in Eq(6). Please use another letter here if not the same, otherwise highly mismatching.  Please use different letters for different quantities to not mismatching the readers. Here and elsewhere below.
%R.: THERE IS NMO NEEED TO CHANGE THE VARIABLE NAME.
is given {by %the following 
\cite{Armesto:2002ny}} %: 
\be
\frac{d\sigma_{q\bar{q}}^A}{d^2\vec{b}} &=& 2\,\left[1-
\exp\left(-\frac{1}{2} \,AT_A(b)\,\hat \sigma(x, \vec{r})\right)
\right],     
\label{GG-nucDIP}
\ee
where {the quantity $A$ denotes the atomic mass number and} $\sigma(x, \vec{r})$ denotes the dipole--proton cross-section. The underlying physical assumption in 
Equation~(\ref{GG-nucDIP}) is that multiple scattering occurs on individual nucleons, as described by the saturating form in \mbox{Equation~(\ref{IP-satDIP}).}

For nuclear targets, the thickness function {is %will 
 to 
 } 
be 
obtained from the Woods--Saxon distribution~\cite{PhysRev.95.577} for the nuclear density~\cite{DEVRIES1987495}, $\rho_A (\vec{s},z)$, %in the following way:
 {as follows:} 
\be
\label{TBNUCLEI}
T_A(b) & = & \int_{-\infty}^{\infty} dz\, \rho_A(\vec{b},z), \quad \int d^2\vec{b}\,T_A(b) =1,\\
\rho_A(\vec{s},z) & = & \frac{\rho_0}{1+\exp\left(\frac{r-R_A}{d} \right)},\quad r = \sqrt{\vec{s}^2+z^2}
\ee
which is {quite a good}
 approximation for nuclei with $A \geq 4$. {Here, %The 
$\vec{s}$ denotes the transverse distance relative to $z$ direction, %MDPI: please add the definition in wording. 
%R.: DEFINITION ADDED.
the }
parameter $\rho_0$ denotes the density at the center of the nucleus, and, for heavy nuclei, the nuclear radius $R_A \simeq 1.12A^{1/3}$ fm and the skin depth $d=0.54$ fm are commonly taken from high-energy electron scattering measurements~\cite{DEVRIES1987495}. 

{As} %MDPI: the paragraph prtly rewritten to fit better. Please confirm the intended meaning retained.  
% R.: YES, WE AGREE. 
 a final remark on the theoretical approach considered in the present {study%work
, to say is that recently, it} %there
 has been %recent 
 {a considerable%important
 } 
 progress {reached} with {the} NLO calculations within the dipole framework. It {%would 
 might %MDPI: please confirm the intended meaning is retained. 
 % R.: YES, WE AGREE. 
 be 
 }
 timely to discuss how {those findings affect} %MDPI: please confirm the intended meaning is retained. The "it" is ambiguous, uncler. 
 % R.: YES, WE AGREE. 
 %it affects 
 {the %presented 
 results
 presented here.}  
By comparing the phenomenology %for 
{of} vector mesons {production}  
with {the} NLO CGC {calculations}~\cite{Mantysaari:2022kdm,Roa:2023skv}{, one finds}%, %there are 
 ~significant deviations 
 only in some regions of the phase space 
(e.g., {at relatively} low photon-nucleon center-of-mass energy $W$ %MDPI: 1. Please define W. 2it would be very useful to provide the number of W: "below X", instead of "relatively". Please consider. 
% YES, W HAS BEEN DEFINED.
for fixed $Q^2$). %It 
{However, it}  
 is %hard %MDPI: colloquial, not recommented for scientific papers. Please the consider replacing. Here and elsewhere below. 
 % R.: YES, WE AGREE. 
 {quite a complicated task} 
to distinguish {between} the {LO and %result from 
NLO results}, given the data accuracy for {vector meson} production. Another 
 {essential} 
%important 
 point is that the {NLO} predictions {are} available %in NLO  %are 
only for protons, and no numerical results are available %yet 
for nuclear targets {so far}.

In {Section \ref{sec:3} just below,} %the next section, 
we apply the obtained cross-section for vector meson photoproduction in $\gamma A$ scattering to the corresponding coherent production in {$AA$} %nucleus--nucleus 
collisions. The associated rapidity distributions are also investigated.

% \begin{figure}[H]
%   \centering
%   \subfigure[]{\includegraphics[width=0.45\textwidth]{figs/pp/RapDist_pp_IPSAT-crop.pdf}} 
%   \subfigure[]{\includegraphics[width=0.45\textwidth]{figs/pp/RapDist_pPsi2Sp_IPSAT-crop.pdf}} 
% \caption{(a) The rapidity distribution for $J/\psi$ photoproduction in proton--proton collisions at 13 TeV. The solid line represents the predictions using boosted Gaussian wavefunction and our prediction by using IPSAT model. (b) The rapidity distribution for $\psi (2S)$ at the same energy. Data from LHCb Collaboration \cite{LHCb:2018rcm}. }
% \label{fig:3}
% \end{figure}

\section{Results and discussions}
\label{sec:3}

Let us start by calculating the rapidity distribution for coherent vector meson production in proton--proton collisions. The equivalent photon approximation 
(EPA) {is %will  
to} be 
used. Thus, the production cross-section in hadron--hadron collisions {%will   be 
 is} factorized as a convolution of photon flux, {$dN_{\gamma}^p/d\omega$, } %MDPI; is \omega of the same meaning as in Eq.(21)? If not please use different letter  here/there. 
 %R.: YES, SAME MEANING.
and the photoproduction cross-section, $\sigma (\gamma p \rightarrow Vp)$. The rapidity 
distribution of the process $pp \rightarrow pVp$ is given {by}% %MDPI: 1. the subscript "gap" moved into the regular (Roman/noral) fon as is a word, not a product of variables or an index. Here and elsewhere below. 2. The comma added at the end. Same for other eqs.Please consider 
% R.: YES, WE AGREE. 
  ~the following~\cite{Bertulani:1987tz,Baur:2001jj,Bertulani:2005ru,Baltz:2007kq,Klein:2020fmr}: 

%\centering %% If there is a figure in wide page, please release command \centering
% R.: YES, WE AGREE. 
\begin{equation}
\frac{d \sigma^{pp}}{dy}= 
\left[{\cal{S}}^{2}_{\rm gap} (\omega^{-})\omega^{-} \, \frac{dN_\gamma^p(\omega^{-})}{d\omega}\, \sigma^{\gamma p
\rightarrow Vp}(\omega^{-})\right] 
+\left[{\cal{S}}^{2}_{\rm gap} (\omega^{+})\omega^{+} \, \frac{dN_\gamma^p(\omega^{+})}{d\omega}\, \sigma^{\gamma p
\rightarrow Vp}(\omega^{+})\right]%
, 
\label{eq:excVMpp}
\end{equation}

where $\omega^{\pm}\approx (M_V/2)e^{\pm y}$ accounts for the source-target symmetry appearing in $pp$ collisions, where each proton simultaneously acts as a source and target of photons. 
%The rapidity gap survival factor is denoted by ${\cal{S}}^{2}_{\rm gap} (\omega)$ on $pp$ collisions. 
{${\cal{S}}^{2}_{\rm gap} (\omega)$ denotes the rapidity gap survival factor in} %MDPI: fits better. Please consider
% R.: YES, WE AGREE. 
 $pp$ collisions. 
In our numerical calculations, the photon flux from a proton is given by the dipole formula for the electric form 
{factor} %%MDPI: We have adjusted the order of the references, please confirm the change.
% R.: YES, WE AGREE WITH THE CHANGE.
~\cite{Drees:1988pp}: 
\be
\frac{dN_\gamma^p(\omega)}{d\omega} &=& \frac{\alpha}{2\pi \omega} 
\left[1+ \left(1-\frac{2\omega}{\sqrt{s_{_{NN}}}}
\right)^2\right] \left(\ln{\Omega} - \frac{11}{6} + \frac{3}{\Omega} - \frac{3}{2\Omega^2}
+\frac{1}{3\Omega^3}\right) \, \, ,
\label{eq:fluxpp}
\ee
{where $s_{_{NN}}$ is the energy squared in the nucleon-nucleon center-of-mass (c.m.),} \linebreak   %MDPI: added as not defined before. 
 $\Omega = 1 + 0.71 \, {\rm GeV}^2/Q_{\rm min}^2$ %MDPI: the formula is rewritten, please confirm.
 % R.: YES, WE AGREE. 
 and the {minimum momentum 
transferred } %MDPI: Q was called virtuaity all the way before. Please check it is the same meaning variable here, otherwise please use another than Q letter here. 
% R.: YES, SAME MEANING.
$Q_{\rm min}^2 = \omega^2/[\gamma^2(1-2\omega/\sqrt{s_{_{NN}}})]\approx \omega^2/\gamma^2$, %( $\sqrt{s_{_{NN}}}$ denotes the hadron--hadron {c.m.} %center-of-mass 
energy). 
{where here $\gamma = %(
 \sqrt{s_{_{NN}}}/2m_N%) %MDPI: no parentheses necessary, clear enough. 
 % R.: YES, WE AGREE. 
 $ represents} 
  the Lorentz factor of a single beam. %MDPI: 1. moved just after use. 2. Looks the gamma here is NOT the same quantity as one used in eq(2). As so, please use different than gamma letter in Eq(2) or here. Please also avoid using gamma as mismatching with the photon nomenclature.
  % R.: NO PROBLEM WITH NOTATION. IT IS UNDERSTANDABLE FOR RESEARCHERS IN THE TOPIC.
 {Then, the} $\gamma p$ c.m. energy is $W_{\gamma p} \simeq 
 \sqrt{2\omega  \sqrt{s_{_{NN}}} }$.  
 %\sqrt{s_{_{NN}}})^{1/2}$.  %MDPI: rewritten, makes it clearer, please consider. 
%and $\gamma = (\sqrt{s_{_{NN}}}/2m_N)$ denotes the Lorentz factor of a single beam.
% R.: YES, WE AGREE. 

\begin{figure}[H]
\centering
\begin{tabular}{c c}
\includegraphics[scale=0.47]{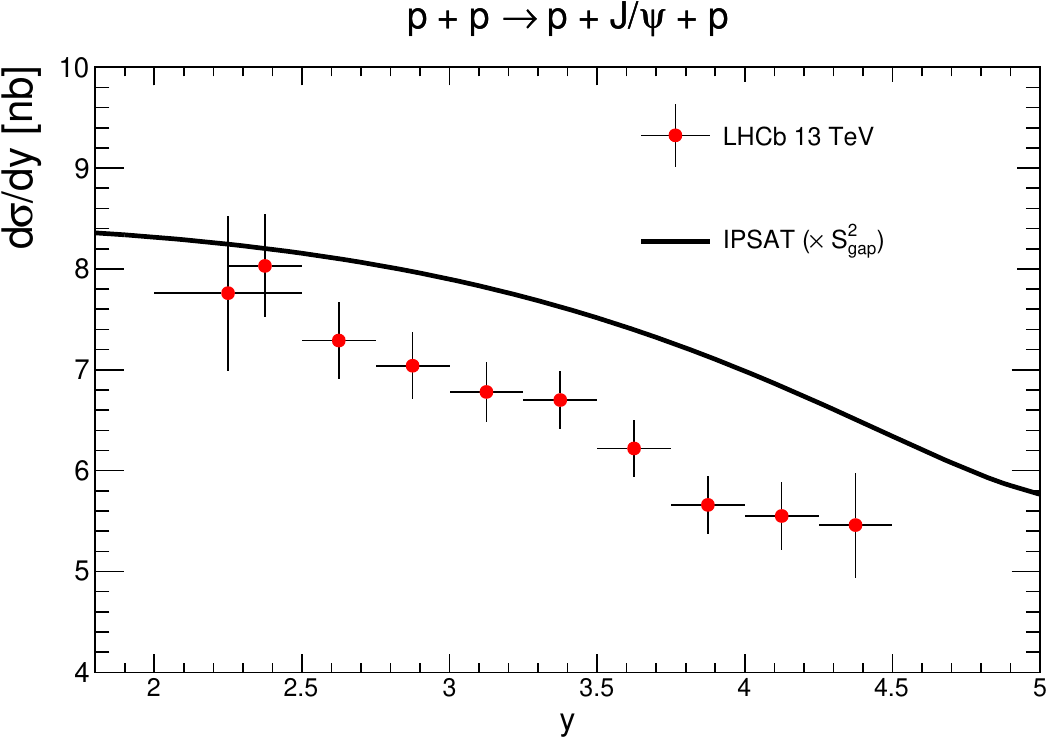}& 
\includegraphics[scale=0.47]{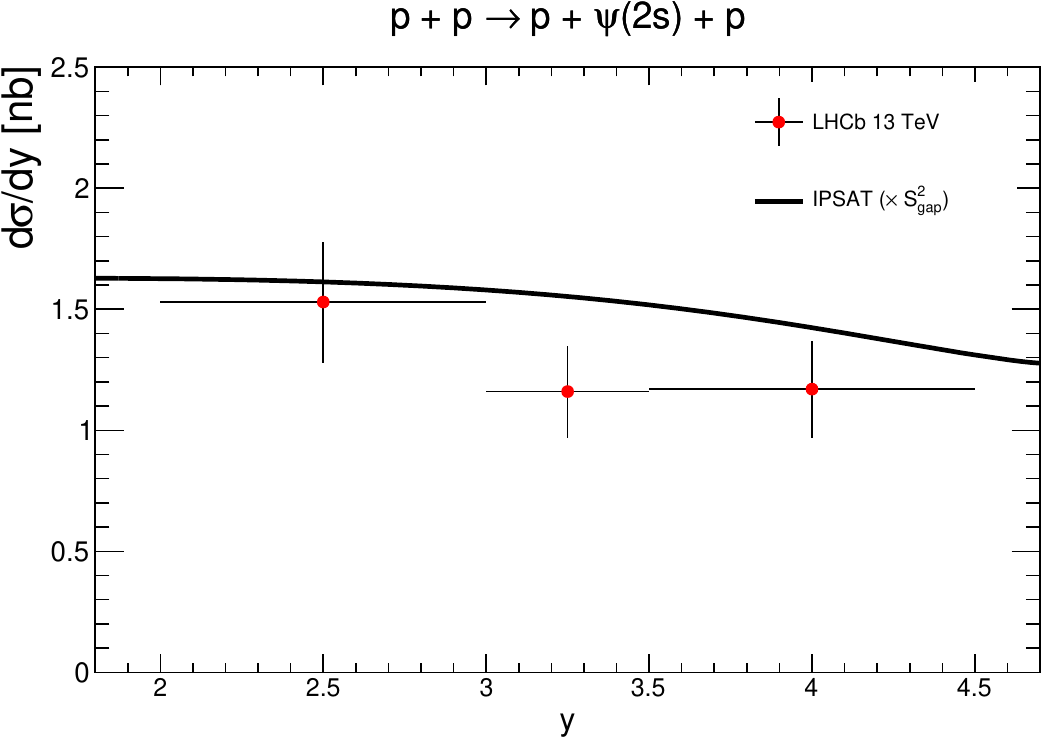} \\
(\textbf{a}) & (\textbf{b}) 
\end{tabular}

 \caption{%(\textbf{a}) The rapidity distribution for $J/\psi$ photoproduction in proton--proton collisions at 13 TeV. The solid line represents the predictions using the 
 %boosted Gaussian wavefunction, and our prediction using the IPSAT model. (\textbf{b}) The rapidity distribution for $\psi (2S)$ at the same energy. Data from the LHCb 
 %collaboration~\cite{LHCb:2018rcm}. }
 {Comparison} %MDPI: We moved the all figures to the end of the first mention. 2. The caption is rewritte, simplified, clarified. Please confirm the revisions.
 % R.: YES, WE AGREE. 
 of the model (solid line) and the data (dots) for the rapidity distribution for $J/\psi$  (\textbf{a}) and $\psi (2S)$ (\textbf{b}) photoproduction in 
 proton--proton collisions at the collision center-of-mass (c.m.) energy of 13 TeV.  The model represents 
 {our prediction using the IPSAT model %MDPI: there is only one line, so only one type of the prediction. Please check.
 % R.: YES, WE AGREE. 
 with the boosted Gaussian wavefunction 
 (see text for details).} %MDPI: clarfies where to find the model description: here, not elsewhere beyond, and also no need for description of the legend. 
 % R.: YES, WE AGREE. 
 The data are the LHCb Collaboration measurements~\cite{LHCb:2018rcm}. 
 }
\label{fig:1}
\end{figure}

In Figure~\ref{fig:1}, the results for $J/\psi$ and $\psi (2S)$ photoproduction in $pp$ %proton--proton 
collisions at {the collision c.m. energy $\sqrt{s}=13$~TeV} (with {the} data {by} %from 
 the LHCb {Collaboration} %MDPI: a capital "C" for "Collaboration". Here and elsewhere below.  
 % R.: YES, WE AGREE. 
\cite{LHCb:2018rcm}) are presented, taking into account the boosted Gaussian wavefunction and our results for 
the IPSAT model (solid line). The rapidity gap survival factor ${\cal{S}}^{2}_{\rm gap}$ is taken from Ref.~\cite{Jones:2016icr}. In Figure~\ref{fig:1}a, the shape of the 
distribution for $J/\psi$ {overestimates} %does not fit 
 the data  %well 
 at forward rapidities. However, the overall normalization overestimates the data mostly at very  
 forward rapidities in this case. The reason can be traced back to the \linebreak  $\gamma p\rightarrow J/\psi+p $ cross-section within the same model, which overestimates the 
extracted high-energy LHCb data~\cite{Fagundes:2022bzw}, while accurately describing the data points at low and intermediate energies. Regarding the $\psi (2S) $ meson 
photoproduction, as shown in Figure~\ref{fig:1}b, the calculation is straightforward as the wavefunctions for excited states are {%well 
known} in the boosted Gaussian parametrization (see Equations \eqref{eq:phi1s} and \eqref{eq:phi2s}). In this case, the data description is improved with respect to the bound~state. 

% Discussion pA
 %With 
{For} $pA$ %proton--nucleus 
 collisions, following the convention that the nucleon is incident from the right and the nucleus from the left,   
the rapidity distribution of 
process $pA \rightarrow pVA$ {reads} %is written as follows: 
\be
\frac{d \sigma^{pA}}{dy}=
\left[\omega^- \, \frac{dN_\gamma^A(\omega^-)}{d\omega}\, \sigma^{\gamma p
\rightarrow Vp}(\omega^-)\right]
+ \left[\omega^+ \, \frac{dN_\gamma^p(\omega^+)}{d\omega}\, \sigma^{\gamma A
\rightarrow VA}(\omega^+)\right] 
,
\label{eq:excVMpA}
\ee
where $\omega^-$ and $\omega^+$ hl{address} %denote 
photons from the nucleus and proton, respectively. Here, a reliable analytic approximation for ultraperipheral $AB$ collisions is given by the photon flux 
integrated over radii larger than {that covered by the nuclei,} %MDPI: defines, otherwise quite unclear. 
 $R_A + R_B$. 
The numerical result, considering an extended nucleus (described by the nuclear form factor), yields a harder photon spectrum for heavy nuclei at the same photon energy. 
Regarding $pA$ collisions, the flux from the nucleus with charge $Z$ and atomic mass number $A$ is evaluated analytically and given {by} %the following
~\cite{Bertulani:1987tz} %: 
\begin{eqnarray}
\frac{dN_\gamma^A(\omega)}{d\omega} =
\frac{2Z^2 \alpha}{\pi \omega} \left[ \xi^{pA}K_0(\xi^{pA})
K_1(\xi^{pA}) 
- \frac{(\xi^{pA})^2}{2} \left(K_1^2(\xi^{pA})-K_0^2(\xi^{pA})
\right) \right] \, \, , 
\label{eq:fluxpA}
\end{eqnarray}
where the parameter $\xi^{pA}$ is given by
$\xi^{pA} = \omega(R_p + R_A)/\gamma$, {with %where 
$R_p$ %is 
the } 
effective radius of the proton. For the photon flux in $AA$ collisions, in Equation~(\ref{eq:fluxpA}), %we 
{it is to} replace {$\xi^{pA}\rightarrow \xi^{AA}%$ with $\xi^{AA} %MDPI: shorten, fits better, no extra wording.
= \omega(2R_A)/\gamma$}.

\begin{figure}[H]

\centering %% If there is a figure in wide page, please release command \centering
% R.: YES, WE AGREE. 
\begin{tabular}{c c}
\includegraphics[scale=0.5]{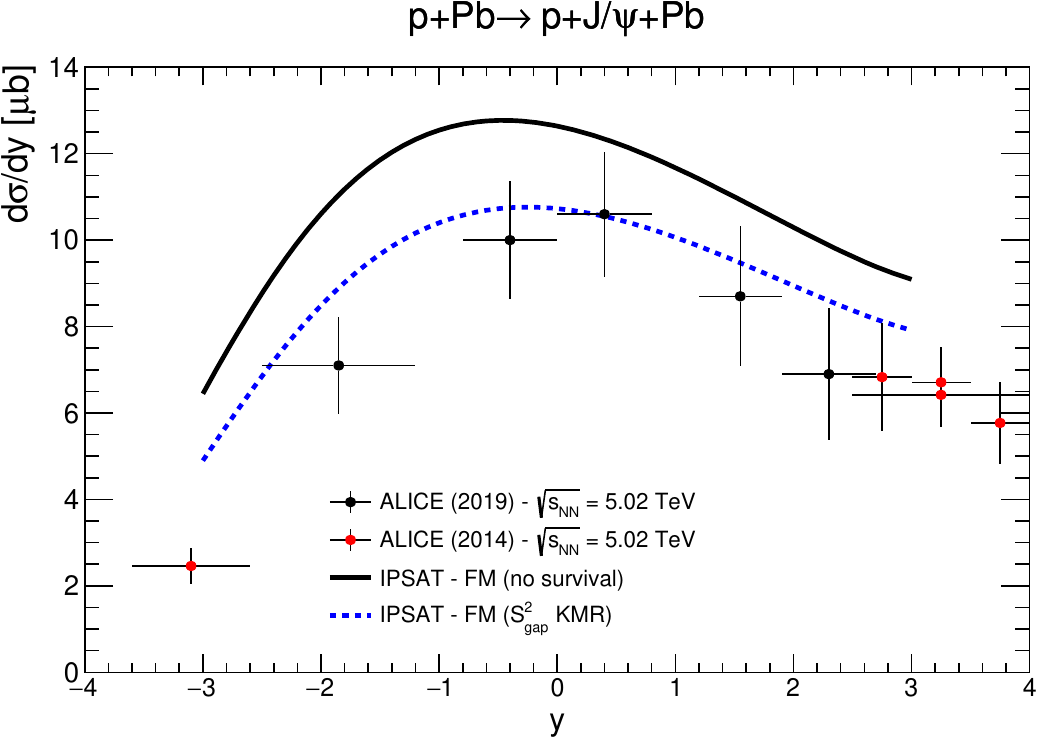}&
 \includegraphics[scale=0.5]{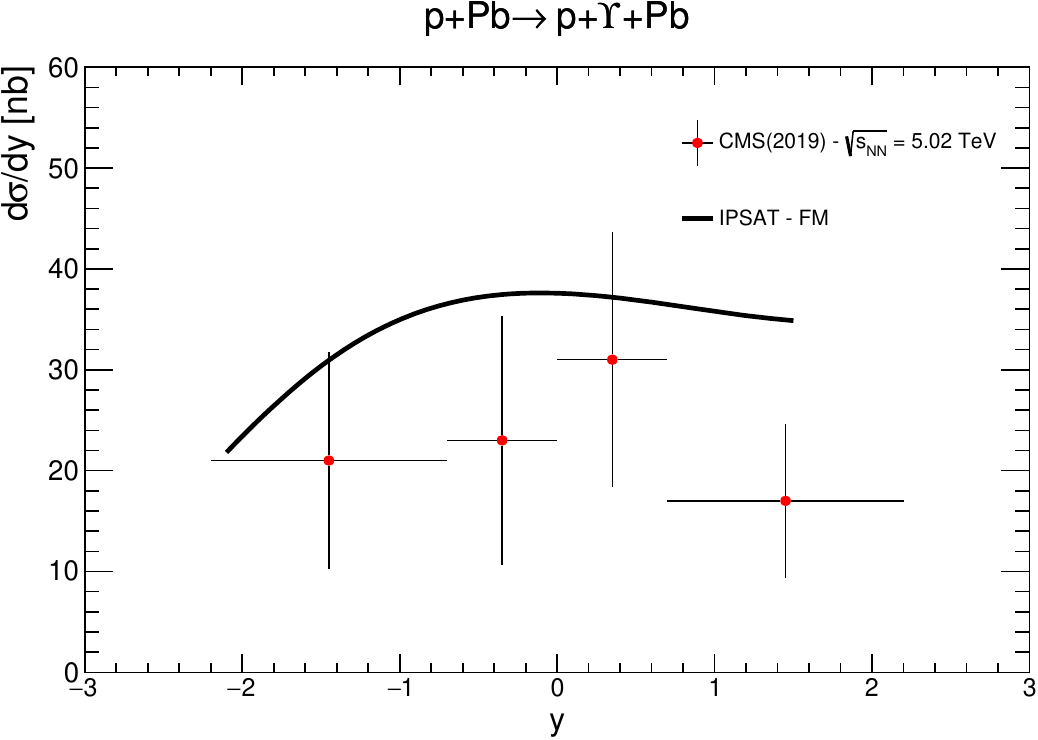} \\
(\textbf{a}) & (\textbf{b}) 
\end{tabular}

   \caption{%(\textbf{a}) 
%The rapidity distribution for $J/\psi$ coherent production in pPb collisions at 5.02 TeV. Results of the boosted Gaussian and the IPSAT model are presented. Data from the ALICE 
%collaboration~\cite{ALICE:2014eof,ALICE:2018oyo}. The solid line represents the prediction without imposing rapidity gap survival suppression, whereas the dotted line represents 
%its inclusion. (\textbf{b}) The rapidity distribution for $\Upsilon$ coherent production in pPb collisions at 5.02 TeV. Data from the CMS collaboration~\cite{CMS:2018bbk}.}
  {Comparison} %MDPI: We moved the all figures to the end of the first mention. 2. The caption is rewritten, simplified, clarified. Please confirm the revisions.
  % R.: YES, WE AGREE.  CONFIRMED!
 of the models (lines) and the data (markers) for the rapidity distribution for $J/\psi$  (\textbf{a}) and $\Upsilon$ (\textbf{b}) photoproduction in 
 proton--lead collisions at the nucleon--nucleon c.m. energy of 5.02 TeV.  The model represents 
 {our predictions using the IPSAT model %MDPI: there is only one line, so only one type of the prediction. Please check.
 with the boosted Gaussian wavefunction implemented,  
 without (solid line) and with (dashed line) inclusion of rapidity gap survival suppression  
 (see text for details).} %MDPI: clarfies,  then no need for description of the legend. 
 % R.: YES, WE AGREE. 
 The data are the measurements of the ALICE \cite{ALICE:2014eof,ALICE:2018oyo} and CMS  \cite{CMS:2018bbk}~Collaborations, as indicated.
 }
 \label{fig:2}
\end{figure}

\begin{figure}[H]
\centering
\begin{tabular}{c c}
\includegraphics[scale=0.48]{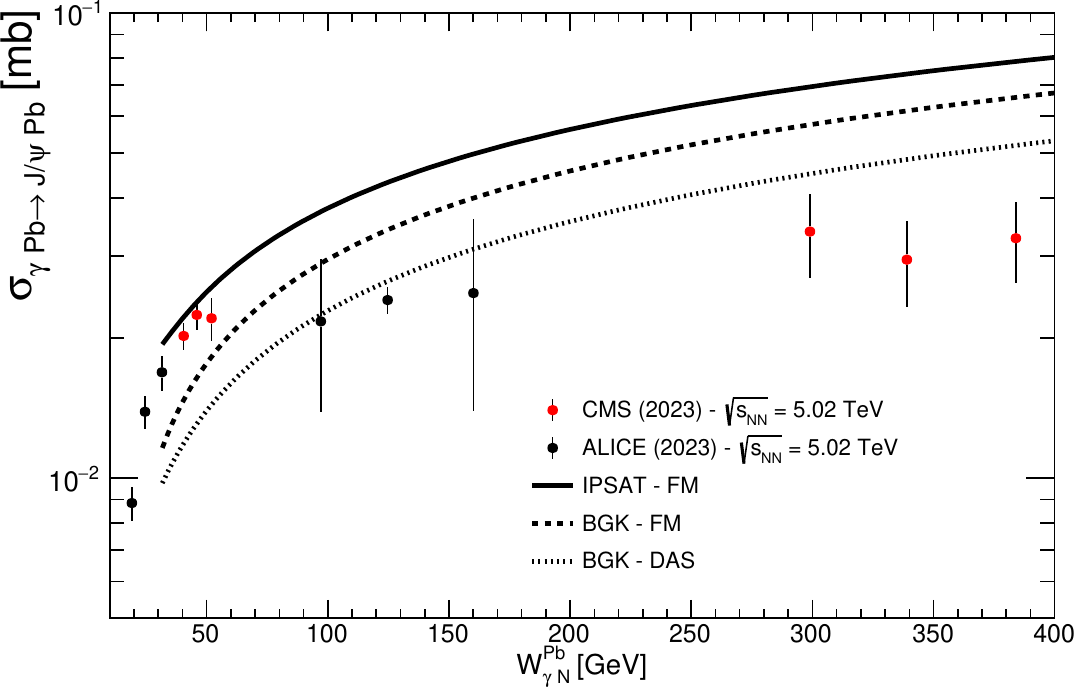}& 
\includegraphics[scale=0.48]{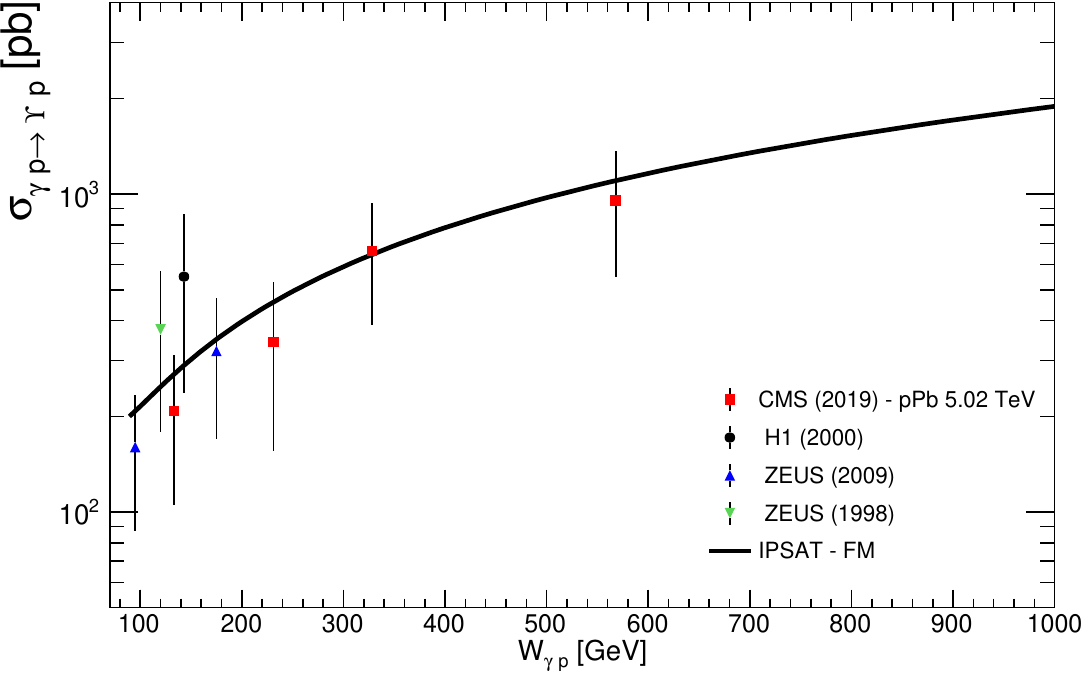} \\
(\textbf{a}) & (\textbf{b}) 
\end{tabular}

   \caption{   {Comparison} 
% R.: YES, WE AGREE. CONFIRMED!
 of the models (lines) and the data (markers) for the photonuclear cross-section $\gamma+Pb\rightarrow J/\psi+Pb$ (\textbf{a}) and $\gamma+Pb\rightarrow 
\Upsilon +p$
 (\textbf{b})
as a function of the photon--nucleon c.m. energy, $W_{\gamma N}$. 
 The models represent the  
 {predictions using the IPSAT (solid line),  BGK (dashed line), and  the BGK-DAS (dotted line) models 
 (see text for details).} 
 %MDPI: clarfies,  then no need for description of the legend. 
 % R.: YES, WE AGREE. 
 The data are the measurements by the CMS Collaboration in PbPb collisions~\cite{CMS:2023snh} 
and the ALICE Collaboration in $p$Pb collisions \cite{ALICE:2014eof,ALICE:2018oyo} at the nucleon-nucleon c.m. energy of 5.02 TeV %MDPI: please add the ref. 
(\textbf{a}) and of the CMS Collaboration in $p$Pb collisions  at the nucleon-nucleon c.m. energy of 5.02 TeV \cite{CMS:2018bbk} and 
the H1 \cite{H1:2000kis} and ZEUS \cite{ZEUS:1998cdr,ZEUS:2009asc}
Collaborations in $ep$ collisions at  the photon--proton c.m. energy $26 <W< 300$ GeV 
%R.: REFERENCEES ADDED AT THE END OF REFERENCE SECTION. PLEASE, CONVERT TO JOUNAL FORMAT. 
(\textbf{b}), as indicated. 
}
  \label{fig:3}
\end{figure}

Let us now move to coherent meson production in ultraperipheral proton--nucleus collisions. Theoretical predictions are compared to 
 {the} ALICE {Collaboration %MDPI: looks jargon otherwise. Same just below.  
 % R.: YES, WE AGREE. 
 data on} charmonium production~\cite{ALICE:2014eof,ALICE:2018oyo} %MDPI: the ref. displaced, fits better, clarifies.  Same just below.
 and {the} %from 
 CMS {data on} %for
 bottomonium production~{Collaboration}~\cite{CMS:2018bbk} %, 
 %in both cases 
 {in $p$Pb collisions} at $\sqrt{s_{NN}}=$~5.02 TeV. 
 %In 
 {Figure~\ref{fig:2}a presents}%, we present 
 the {rapidity distribution for} %results using 
 the $J/\psi$ production, {whereas %in 
 Figure~\ref{fig:2}b shows %, 
 the} corresponding {distribution %is shown 
 for $\Upsilon$}  production. 
 The {role of %played by 
 the} rapidity gap survival factor in $p$A %proton--nucleus 
 collisions is investigated for the $J/\psi$ case. The average ${\cal{S}}^2_{\rm gap}$ 
is taken from Ref.~\cite{Flett:2022ues}. 
 %In Figure~\ref{fig:2}a, the result without this rapidity interval is labeled by the solid curve, and its inclusion is represented with the dotted curve. %MDPI: is described in and is clear from the fugure, not necessary, rather mismatching, repeated in the text. Please consider.
 % R.: YES, WE AGREE. 
 The main effect of {including} this rapidity interval is the overall normalization. The prediction without the survival {factor %(solid line) 
 is} also {given} %presented  
 for the $\Upsilon$  photoproduction in {Figure} \ref{fig:2}b. %It
 {The prediction} %MDPI; ;otherwise ambiguous, can be treated as photoproduction but is the model as it looks. Please confirm thr intended meaning os retained. 
 % R.: YES, WE AGREE. MEANING RETAINED.
 is computed by using the $\gamma+p\rightarrow \Upsilon +p$ cross-section shown in 
 Figure~\ref{fig:3}b. The data description in this case is {exceptionally %very
 good}. The inclusion of absorption corrections in ultraperipheral $p$Pb collisions {is obtained to describe the} %describes the existing %MDPI: extra wording, clear you compare to the existing data. 
 data in a more satisfactory way. 

% Discussion AA
Finally, in {$AA$} %nucleus--nucleus 
collisions, the rapidity distribution of the process $AA \rightarrow AVA$ {reads}% as follows
:
\be
\frac{d \sigma^{AA}}{dy}= 
\left[\omega^-\, \frac{dN_\gamma^A(\omega^-)}{d\omega}\, \sigma^{\gamma A
\rightarrow VA}(\omega^-)\right]
+\left[\omega^+ \, \frac{dN_\gamma^A(\omega^+)}{d\omega}\, \sigma^{\gamma A
\rightarrow VA}(\omega^+)\right]
, 
\label{eq:excVM}
\ee
with $\omega^-$ and $\omega^+$ just %simply 
denoting equivalent photons from the nucleus incident from the left and right, respectively. 
The flux from the nucleus with {the} charge $Z$ and atomic mass number $A$ is evaluated analytically and {is} 
given by~\cite{Bertulani:1987tz} %:

\begin{eqnarray}
\frac{dN_\gamma^A(\omega)}{d\omega} =
\frac{2Z^{2 \alpha}}{\pi \omega} 
\left[ \xi K_0(\xi)
K_1(\xi) 
- \frac{(\xi)^2}{2} \left(K_1^2(\xi)-K_0^2(\xi)
\right) \right] \, \, , 
\label{eq:fluxpA}
\end{eqnarray}
{where} $\xi=2\omega R_{A}/\gamma$. 
%, $\gamma = (\sqrt{s_{_{NN}}}/2m_N)$ denotes the Lorentz factor of a single beamand $\omega^{\pm}\approx (M_V/2)e^{\pm y}$. %MDPI: already defined above.
% R.: YES, WE AGREE. 

\vspace{-8pt}

\begin{figure}[H]

\centering
\begin{tabular}{c c}
\includegraphics[scale=0.5]{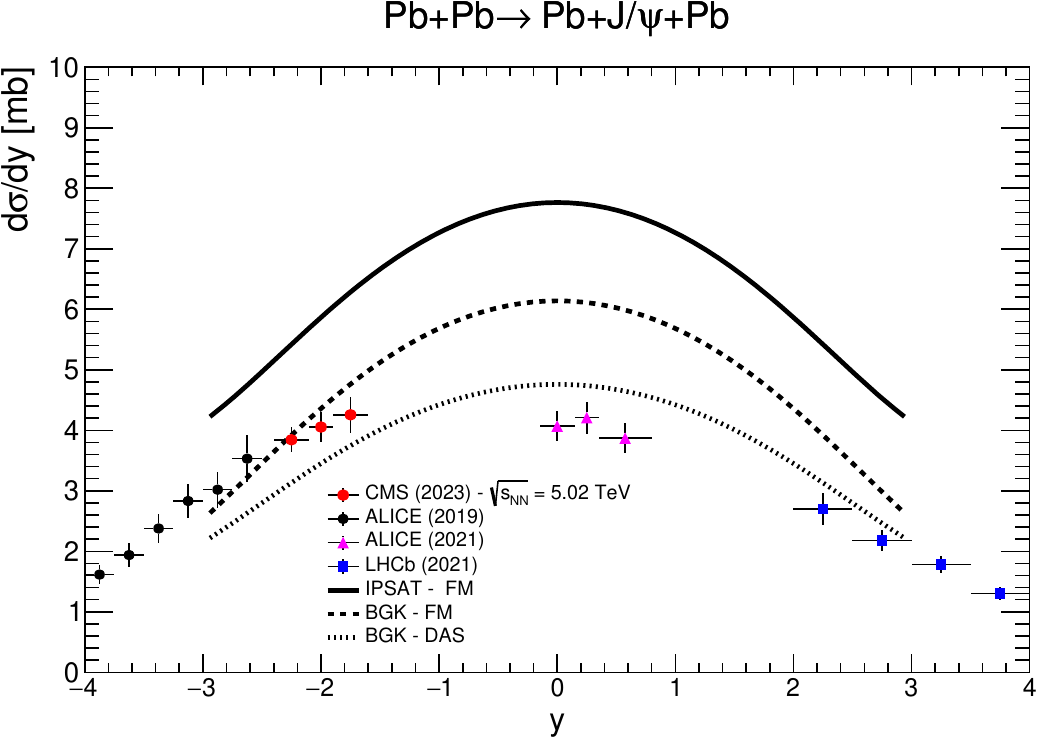}&
 \includegraphics[scale=0.5]{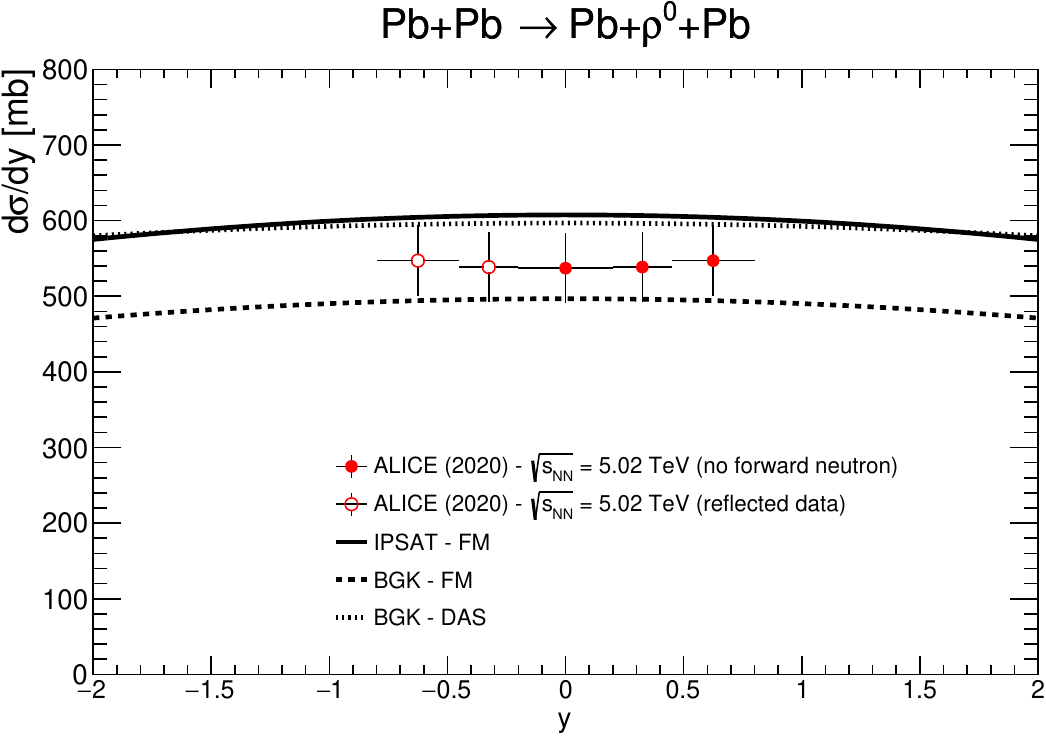} \\
(\textbf{a}) & (\textbf{b}) 
\end{tabular}

   \caption{
%(\textbf{a}) Rapidity dependence for coherent $J/\psi$ production in PbPb collisions at 5.02 TeV. The same notation as in the previous plot. Data from 
%ALICE~\cite{ALICE:2019tqa,ALICE:2021gpt}, LHCb~\cite{Bursche:2018eni}, and CMS~\cite{CMS:2023snh} collaborations. (\textbf{b})~Rapidity dependence for 
%coherent $\rho^0$ production in PbPb collisions at 5.02 TeV compared to ALICE data~\cite{ALICE:2020ugp}.
 {Comparison} of the models (lines) and data (markers) for the rapidity distributions of coherent $J/\psi$ (\textbf{a})  and $\rho^0$ (\textbf{b}) production 
in PbPb 
collisions at the nucleon--nucleon c.m. energy of 5.02 GeV.   
  The models represent the {predictions using the IPSAT (solid line), BGK (dashed line), and  the BGK-DAS (dotted line) models 
 (see text for details).} %MDPI: clarfies,  then no need for description of the legend. 
 The data are the measurements by the ALICE~\cite{ALICE:2019tqa,ALICE:2021gpt}, LHCb~\cite{Bursche:2018eni}, and CMS~\cite{CMS:2023snh} Collaborations  
(\textbf{a}) and by the ALICE Collaboration~\cite{ALICE:2020ugp} (\textbf{b}), as indicated. 
}
  \label{fig:4}
\end{figure}

\begin{figure}[H]
\centering
\includegraphics[scale=0.6]{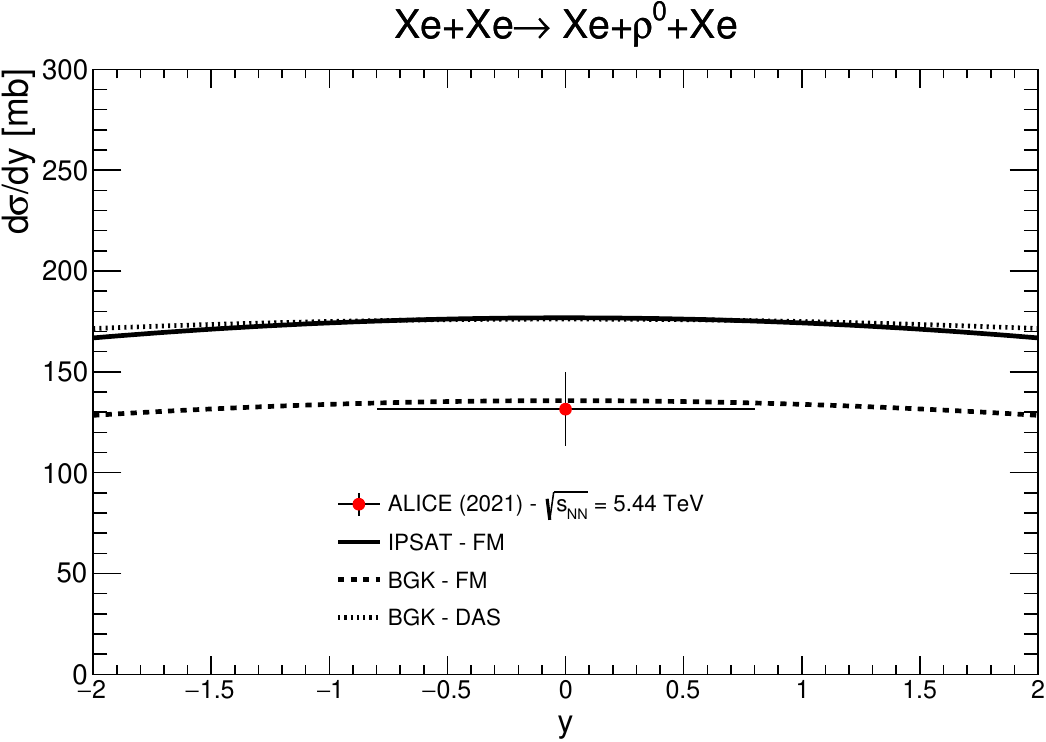}
 \caption{%Rapidity distribution for the coherent photoproduction of $\rho^0$ mesons in XeXe UPC in midrapidity at 5.44 TeV. 
%collaboration~\cite{ALICE:2021jnv}.
{Comparison} of the models (lines) and data (dot) for the rapidity distribution of the coherent photoproduction of $\rho^0$ mesons in XeXe UPC at 
midrapidity at the nucleon--nucleon c.m. energy of 5.44 TeV. 
Predictions of the {IPSAT %model 
(solid line), %the 
BGK %model 
(dashed line), and the BGK-DAS %model 
(dotted line) models} are compared with {the} data {by} %from 
the ALICE {Collaboration}~\cite{ALICE:2021jnv}.  
}
\label{fig:5}
\end{figure}

Now, we {address} %investigate
 the photonuclear production of vector mesons in $AA$ %nucleus--nucleus 
collisions at colliders. First, {let us} %we 
consider PbPb and XeXe collisions at LHC at the $NN$ %nucleon 
 c.m. energies {of} 5.02 
TeV and 5.44~TeV, respectively. First, {let us} %we 
focus on the coherent production without nuclear breakup. In Figure~\ref{fig:3}a, we present the result concerning the photonuclear cross-section, $\gamma+\mathrm{Pb}\rightarrow J/\psi+\mathrm{Pb}$, as a function 
{of 
 %the photon--nucleon center-of-mass 
$W_{\gamma N}$, } %MDPI: already defined. 
%R.: YES, WE AGREE.
and considering the boosted Gaussian wavefunction and 
{the %models 
IPSAT% (solid curve)
, 
BGK% (dashed curve)
, 
and BGK-DAS% (dotted curve)
. One finds that the} %The 
IPSAT and BGK models are consistent with {the} 
low-energy data but produce predictions that overestimate {the} 
data~\cite{CMS:2023snh} at %high 
$W_{\gamma N}$ {above 300 GeV}. 
In Figure~\ref{fig:4}a, the results for the coherent $J/\psi$ production, $\mathrm{Pb}+\mathrm{Pb}\rightarrow \mathrm{Pb}+J/\psi+\mathrm{Pb}$, are shown. The rapidity distributions are present in both central and forward 
{rapidity regions.} %$y$. 
The change in the overall normalization follows the results of the photonuclear cross-section, as shown in Figure~\ref{fig:4}a. That is, the models overestimate the data at high 
{c.m.} %MDPI: added as so far you talhed abot gamm-N c.m. energy. Please confirm the addition.
% R.: YES, WE AGREE. 
energies (very 
 forward/backward rapidities). {Certainly,} %Of course, %MDPI: colloquial, highly not recommended for scirntific papers. Please consider the replacing.  
% R.: YES, WE AGREE. 
there is a normalization discrepancy between {the} 
ALICE~\cite{ALICE:2019tqa,ALICE:2021gpt} and LHCb~\cite{Bursche:2018eni} {Collborations} {data what %, %MDPI: no comma.  
%R.: YES, WE AGREE.
%which
} remains a matter of 
discussion. Furthermore, a 
significant suppression in the data relative to theoretical predictions persists at central rapidities. In the literature, within the context of the color dipole picture, this issue can be solved by including contributions from the next Fock states of the virtual photon, $q\bar{q}(ng)$ 
({with} 
$n=1,2,\ldots$ {and $g$ denoting the gluon})~\cite{Luszczak:2021jtr,Henkels:2020qvo,Henkels:2020kju,Ducati:2013bya,Kopeliovich:1999am,Kopeliovich:2022jwe,Kopeliovich:2020has}. These corrections are directly related to the gluon nuclear shadowing phenomenon, as discussed in
Refs.~\cite{Kopeliovich:1999am,Kopeliovich:2022jwe,Kopeliovich:2020has} 
and {in the} references therein. It is worth mentioning that the finite coherence length corrections~\cite{Kopeliovich:2022jwe,Kopeliovich:2020has,Henkels:2020kju} are not 
included in the 
present calculations. {For %the sake of 
completeness,} the analysis for coherent $\rho^0$ production in PbPb collisions is presented in Figure~\ref{fig:4}b %. The notation is the same as in Figure~\ref{fig:4}a, and a 
{in}  comparison to {the} 
ALICE {Collaboration} data~\cite{ALICE:2020ugp}.  
 %  is performed. 
A weaker model-dependence compared to the $J/\psi$ {is} %case has been 
verified. This can be related to the universal behavior of the dipole cross-section, $N(r \gg 1/Q_s) \rightarrow 1 $, for {large} 
 dipoles, which dominate the dynamics for light meson photoproduction. The cross-section for the nucleon target, $\gamma+p\rightarrow \rho^0+p$, was described 
%in a previous publication
{by us earlier in Ref.}  
~\cite{Fagundes:2022bzw}. The coherent $\rho${-}production in XeXe collisions is presented in Figure~\ref{fig:5} and compared %to 
{with the} ALICE {Collaboration} data~\cite{ALICE:2021jnv}. The results 
follow {a %similar 
trend similar to that of} the PbPb collision case.

\vspace{-12pt}

Regarding the predictions for light mesons, some comments are in order. The DAS parametrization (and its modification proposed here) for the gluon density is 
valid in the $x\rightarrow 0$ and $Q^2\rightarrow \infty$ limits. In light {of} meson photoproduction, there is no hard scale, and the validity of the 
approach in this case could be questioned. The answer to this question was already given in 
{our earlier study}%work; see Ref.
~\cite{Fagundes:2022bzw}{, where %. There, 
 we} were able to describe the data from {the proton structure function }  %MDPI: please define the quamntity by wording.
 %R.: YES, NOW DEFINED.
 $F_2$ to {considerably %very 
 low} $Q^2$, the total photoabsorption cross-section, and the $\rho$ photoproduction itself using the dipole{-%+
DAS}-based models. 
 The point is that the %good %MDPI: not necessary, agreement or no agreement says it all. 
 %R.: YES, WE AGREE.
 agreement with the data is not related to the gluon distribution at that {$Q^2$-}%MDPI: please confirm, otherwise the scale is not defined. 
 %R.: YES, WE AGREE.
scale, but rather to the multiple scattering effects embedded in the 
IP-SAT or BGK models. On scales below pQCD validity $\mu \lesssim 1$ GeV, the wavefunction overlap selects large-dipole configurations, for which the cross-section becomes essentially constant in these eikonal-based models, independent of the behavior of the gluon distribution at such {low} 
 scales.

Finally, {let us address} 
the Coulomb nuclear breakup. % is addressed. 
 In this {process,} %case, 
the nuclei participating {undergoes} %in the process undergo 
electromagnetic dissociation (EMD), producing 
neutrons \cite{Baltz:2002pp}. The cross-section for a symmetric ultraperipheral $AA$ collision accompanied by the breakup of one or both nuclei, with the 
production of a vector {meson 
%$V$m, %MDPI: already defined above.
is given by} %the following 
 \cite{Broz:2019kpl,Kryshen:2023bxy}:
\begin{eqnarray}
\sigma (AA\rightarrow V A^{\prime}_iA^{\prime}_j) \propto \int d^2 \vec{b} \int d\omega \frac{d^2 n(\omega,b)}{d\omega d^2\vec{b}} \sigma_{\gamma 
A\rightarrow VA}(\omega) P_{ij}(b)\,\exp \left[ -P_H(b) \right],
\end{eqnarray}
where $P_{ij}(b)=P_i(b)\times P_j(b)$ denotes the probability of nuclear breakup with emissions of $i$ and $j$ neutrons ($i,j=0,1,2,\ldots$), respectively, from the first and second nucleus, and $P_H(b) = \int d^2\vec{r}\,T_A(\vec{r}-\vec{b})[1-\exp(-\sigma_{NN}T_A(\vec{r}))]$ denotes the probability of 
a hadronic interaction { %. The 
 with the inelastic 
 $NN$ %nucleon--nucleon 
cross-section    
at the corresponding c.m. %center-of-mass 
energy,} %is denoted by 
$\sigma_{NN}(\sqrt{s_{NN}})$. In the EPA approximation, the flux of photons, $d^2n/d\omega d^2\vec{b}$, produced with 
{energy $\omega$} %MDPI: please be sure the omega has ALWAYS the SAME meaning, starting Eq.(21)! 
at impact parameter $\vec{b}$, can be expressed by the~following:
\begin{eqnarray}
\frac{d^2 n(\omega,b)}{d\omega d^2\vec{b}} =  \frac{Z^2\alpha}{\pi^2\gamma^2 } \omega \left[k_1^2(\zeta)+\frac{1}{\gamma^2} k_0^2(\zeta) \right], \quad \zeta = \frac{\omega b}{\gamma}.
\end{eqnarray}

\begin{figure}[H]

\centering
\begin{tabular}{c c}
\includegraphics[scale=0.485]{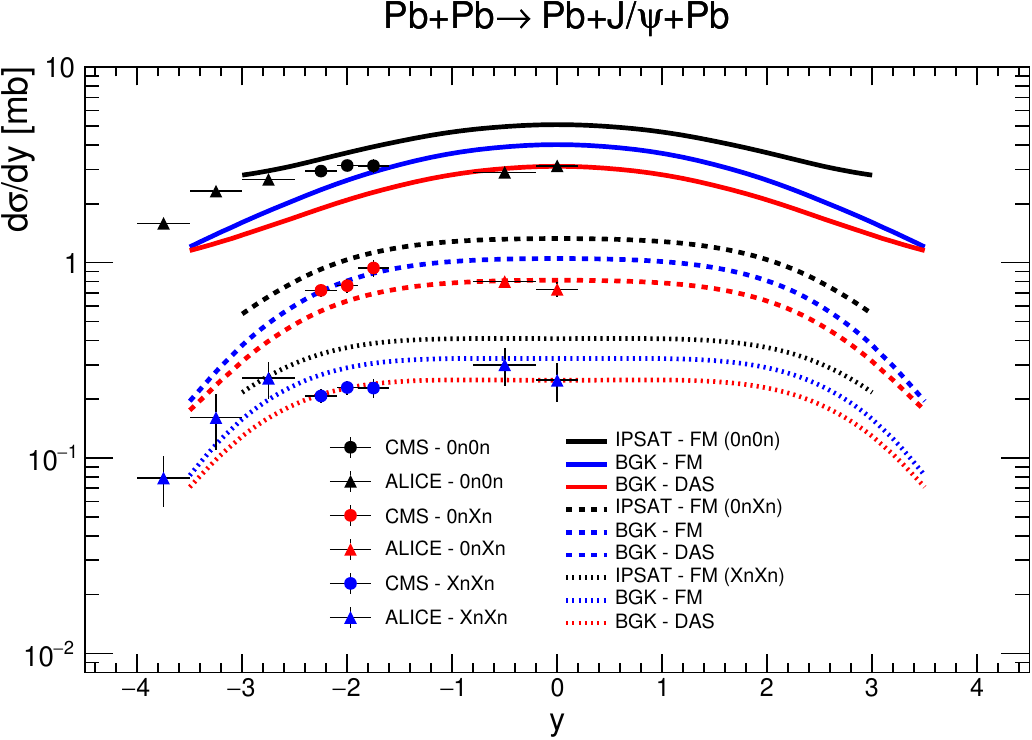}& \includegraphics[scale=0.485]{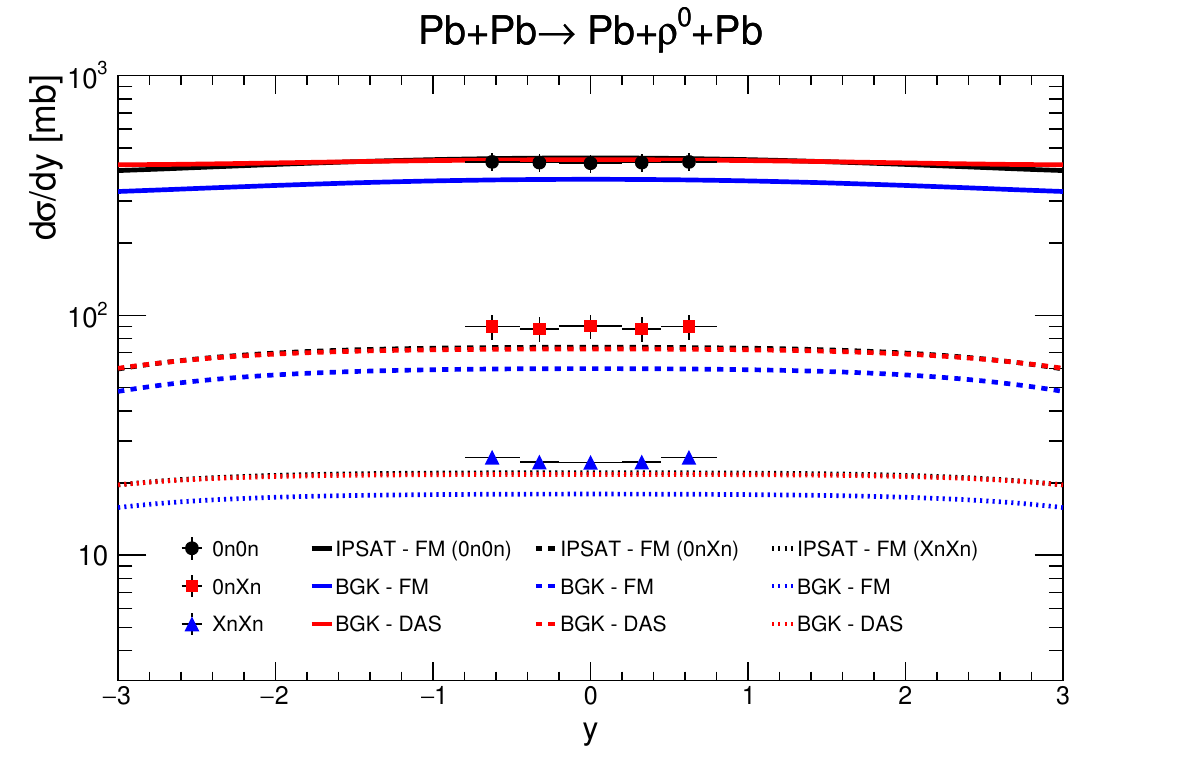} \\
(\textbf{a}) & (\textbf{b}) 
\end{tabular}

   \caption{%(\textbf{a}) Rapidity distribution for coherent $J/\psi$ photoproduction in 0n0n (solid curves), 0nXn (dashed curves) and XnXn (dotted curves) neutron 
%multiplicity classes. Data from the CMS collaboration~\cite{CMS:2023snh} and ALICE collaboration~\cite{ALICE:2023jgu}. Theoretical predictions in each class 
%from the IPSAT (upper), BGK (middle), and BGK-DAS (lower) models. (\textbf{b}) Rapidity distribution for coherent $\rho^0$ photoproduction in the different 
%neutron multiplicity classes using the same notation as in (\textbf{a}). Data from the ALICE collaboration~\cite{ALICE:2020ugp}.
 {Comparison} of the models (lines) and the data (markers) for  the rapidity distribution for coherent $J/\psi$ (\textbf{a}) and $\rho^0$ (\textbf{b}) 
photoproduction in PbPb collisions at the nucleon--nucleon c.m. energy of 5.02 GeV  in three neutron multiplicity classes. Theoretical predictions from the 
IPSAT (upper), BGK (middle), and BGK-DAS (lower) models for 0n0n (solid lines), 0nXn (dashed lines), and XnXn (dotted lines) classeds (see text for details) 
are compared with the measurements by the CMS~\cite{CMS:2023snh} and  ALICE~\cite{ALICE:2023jgu} Collaborations (\textbf{a})  and by the ALICE 
Collaboration~\cite{ALICE:2020ugp} (\textbf{b}), as indicated. 
}
  \label{fig:6}
\end{figure}

In our numerical calculations, in what follows, the Monte Carlo program {\textbf{n$\mathbf{_O^O}$n}} %MDPI: Please confirm if the bold is unnecessary and can be removed. Note that bold is specifically reserved for paper, section titles, years in refs etc. 
 (noon)~\cite{Broz:2019kpl} of forward neutrons for {UPC} %ultra-peripheral collisions 
was used. {The program} %It 
is a ROOT-based code that can be interfaced with existing generators of exclusive vector meson production in UPC or with theoretical calculations of such processes. Setting the notation, $P_{\mathrm{Xn}}$ denotes the probability of the nuclear breakup of one nucleus to a state with any number ($\mathrm{X}$) of neutrons ($\mathrm{n}$). 
 %This 
{ $P_{\mathrm{Xn}}$} %MDPI: clarifies; No "it", "this", "them", "its" etc recommended as mismatching, ambiguous. Please consider. Here and elsewhere in teh paper.
%R.: YES, WE AGREE.
is obtained through the mean number of the Coulomb excitation of the nucleus to any state that emits one or more neutrons, $P_{\mathrm{Xn}}^1$. In Figure~\ref{fig:6}a, the rapidity distributions from coherent $J/\psi$ production in 
{0n0n% (solid curves)
, 0nXn% (dashed curves)
, and XnXn %(dotted curves) 
 neutron} multiplicity classes are presented and contrasted {to the} %against 
experimental measurements {by the} %from 
 CMS~\cite{CMS:2023snh} and ALICE Collaborations~\cite{ALICE:2023jgu}. Theoretical predictions are also given in each class for the IPSAT, BGK, and BGK-DAS 
models. {One finds that at} %At 
large  rapidities, the deviations among the models are quite similar to the predictions without the EMD effect. The BGK model produces better 
{describes the} data 
 %descriptions 
in 
 the 0n0n and 0nXn classes, whereas BGK-DAS provides better {agreed} results for the XnXn class in the case of the CMS 
{experiment} dataset. 
 %In the situation where 
{When} 
both experimental datasets are considered, the overall normalization is 
well {enough}  
described within the theoretical uncertainties (comparing the three implementations of the nuclear dipole amplitude). In effect, given its higher normalization, IPSAT shall be regarded hereafter as an upper bound of the dipole amplitude for $J/\psi$ photoproduction, while the BGK-DAS model 
 %is 
 {as} 
the lower bound. The rapidity distribution for $\rho^0${-}production 
in the same neutron multiplicity classes is shown in Figure~\ref{fig:6}b.  
 %using the same notation for the curves. %MDPI: not necessary, clear from Figure.
 %R.: YES, WE AGREE.
The predictions are compared to {the} measurements by the 
ALICE Collaboration~\cite{ALICE:2020ugp}. As already discussed {above in this Section,} 
in the case without nuclear breakup, the %deviation
 {difference}  
among the models is less pronounced in the 
light{-}meson photoproduction.
 
\begin{figure}[H]
\centering
  \includegraphics[scale=0.6]{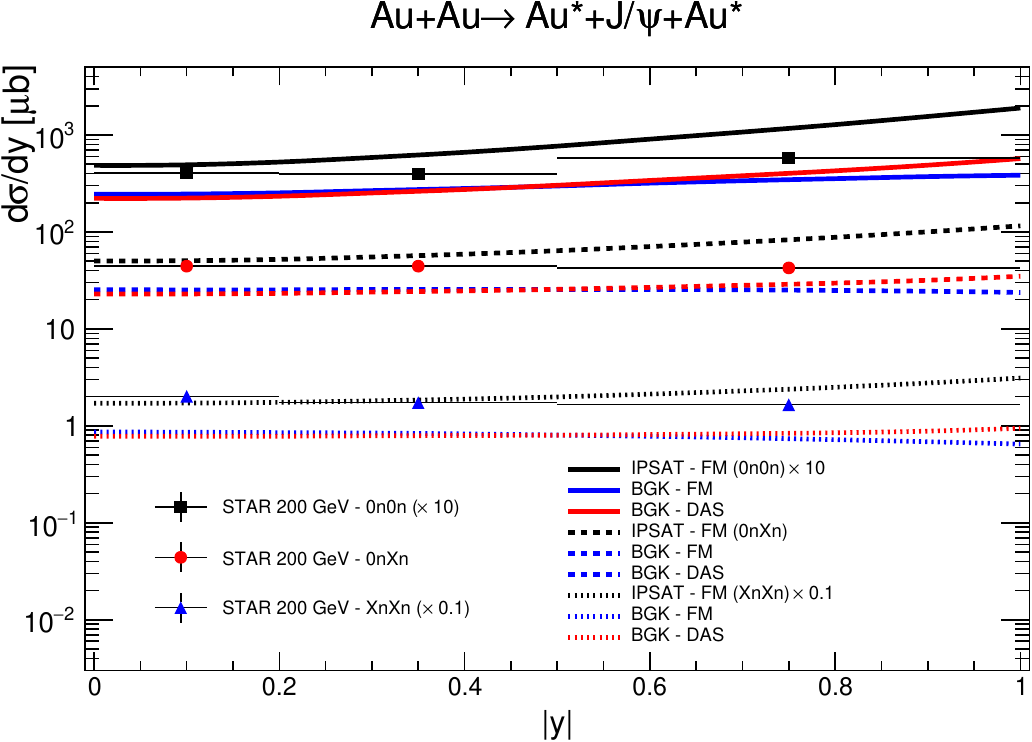}
 \caption{%(\textbf{a}) %MDPI: There is no subfigure label in figure 7, so removed, please check.
 %R.: YES, WE AGREE.
 %Rapidity distribution for coherent $J/\psi$ production in AuAu collisions at 200 GeV. Same notation as in the previous plot. 
 %R.: YES, WE AGREE.
 %Data from STAR~\cite{STAR:2023gpk}.} 
 {Comparison} of the models (lines) and the data (markers) for the rapidity distribution of coherent $J/\psi$ production in AuAu collisions at 200 GeV in 
three neutron multiplicity classes. Theoretical predictions from the IPSAT (upper), BGK (middle), and BGK-DAS (lower) models for 0n0n (solid lines), 0nXn 
(dashed lines), and XnXn (dotted lines) classes (see text for details) are compared with the measurements by the STAR Collaboration~\cite{STAR:2023gpk}, as 
indicated.
}
 \label{fig:7}
\end{figure}

 %Afterward, we address 
 {Lastly, let us consider}  
the {quite%very
 } 
recent RHIC measurements on $J/\psi$ production in AuAu collisions at 200 GeV~\cite{STAR:2023gpk}, for neutron multiplicity classes 0n0n, 0nXn, and XnXn. 
%These 
{The calculations compared to those measurements} %results 
are shown in Figure \ref{fig:7}, which, {to be said} not {quite unexpectedly
 %surprisingly %MDPI: colloquial, not recommended for scientific papers. Please consider the the replacing. here and elsewhere below.
 %R.: YES, WE AGREE.
 }, {demonstrate} %s 
that all data points fall within the uncertainty region previously defined by {the} %following models: 
IPSAT (upper bound) and BGK-DAS (lower bound) {models}.

\vspace{-6pt}

\section{Conclusions and Final Remarks}
\label{sec:conc}

In summary, in this {paper%work}
, the} calculations {have been} %MDPI: fits better or move to the present tense. %were 
 performed considering an analytical gluon distribution based on DAS approximation, leading to predictions for the exclusive vector meson production of 
 $\rho^0$, $J/\psi$, $\psi (2S)$, and $\Upsilon (1S)$. The focus is %was 
 on {the} their 
 coherent production {of the mesons in} %for 
 $pp$, $pA$, and $AA$ collisions at the LHC {and RHIC} energies. 
 This analysis extended the investigation presented {earlier} in Ref.~\cite{Fagundes:2022bzw}, where the models were proposed and compared to structure 
 functions and 
 exclusive meson production in the HERA kinematics. It is shown that the different implementations of the dipole--proton amplitude (IPSAT-FM, BGK-FM, and 
 BGK-DAS) using such a gluon PDF lead to a sizable theoretical uncertainty in the photonuclear cross-section and rapidity distributions. %Of course, 
 {Certainly,} other 
 sources of uncertainties, such as the meson wave function and quark masses in the phenomenological models, {are %were 
 not} addressed at this stage. In numerical 
 calculations, the boosted Gaussian meson wavefunction was used; %one would expect 
 a change in the normalization of the rapidity distributions {is expected} 
 when other parametrizations are considered, such as the light-cone Gaussian (LCG). In general, 
 the rapidity distributions of coherent vector{-}meson production are {obtained to be} 
 fairly {well}  
described by the analytical gluon PDF. The electromagnetic dissociation in 
 nucleus--nucleus collisions has also been studied. Based on the {\textbf{n$\mathbf{_O^O}$n}} %MDPI: Please confirm if the bold is unnecessary  and can be  removed.
 %R.: YES, WE AGREE.
 generator, the rapidity distributions for coherent meson photoproduction in 0n0n, 0nXn, and XnXn neutron multiplicity classes {have been} %were 
computed. Taking into account the theoretical uncertainty, the models {are found to be} %are 
able to describe the recent LHC data for $J/\psi$ and $\rho^0$ production and {the %very 
 recent} RHIC-STAR data.

\begin{acknowledgments}
We thank  Christopher Flett  for discussions on the survival probability in $pA$.  DAF acknowledges the support of the project
INCT-FNA (464898/2014-5), Brazil. MVTM was supported by funding agencies CAPES (Finance Code 001) and CNPq (grant number
303075/2022-8), Brazil.
\end{acknowledgments}

%\reftitle{References}


\begin{thebibliography}{999}

\bibitem[Gribov et~al.(1983)Gribov, Levin, and Ryskin]{Gribov:1983ivg}
Gribov, L.V.; Levin, E.M.; Ryskin, M.G.
\newblock {Semihard processes in QCD}.
\newblock {\em Phys. Rep.} {\bf 1983}, {\em 100},~1--150. [\href{http://doi.org/10.1016/0370-1573(83)90022-4}{CrossRef}]

\bibitem[Mueller and Qiu(1986)]{Mueller:1985wy}
Mueller, A.H.; Qiu, J.W.
\newblock {Gluon recombination and shadowing at small values of $x$}.
\newblock {\em Nucl. Phys. B} {\bf 1986}, {\em 268},~427--452. [\href{http://dx.doi.org/10.1016/0550-3213(86)90164-1}{CrossRef}]

\bibitem[Mueller(2001)]{Mueller:2001fv}
Mueller, A.H.
\newblock {Parton saturation: An overview}.
\newblock In Proceedings of the {Cargese Summer School on QCD Perspectives on
Hot and Dense Matter}, {Cargese, France, 6--18 August} 
2001; pp. 45--72. [\href{http://dx.doi.org/10.48550/arXiv.hep-ph/0111244}{CrossRef}]

\bibitem[McLerran and Venugopalan(1994{\natexlab{a}})]{McLerran:1993ni}
McLerran, L.D.; Venugopalan, R.
\newblock {Computing quark and gluon distribution functions for very large
nuclei}.
\newblock {\em Phys. Rev. D} {\bf 1994}, {\em 49},~2233--2241. [\href{http://dx.doi.org/10.1103/PhysRevD.49.2233}{CrossRef}]

\bibitem[McLerran and Venugopalan(1994{\natexlab{b}})]{McLerran:1993ka}
McLerran, L.D.; Venugopalan, R.
\newblock {Gluon distribution functions for very large nuclei at small
transverse momentum}.
\newblock {\em Phys. Rev. D} {\bf 1994}, {\em 49},~3352--3355. [\href{http://dx.doi.org/10.1103/PhysRevD.49.3352}{CrossRef}]

\bibitem[McLerran and Venugopalan(1994{\natexlab{c}})]{McLerran:1994vd}
McLerran, L.D.; Venugopalan, R.
\newblock {Green's functions in the color field of a large nucleus}.
\newblock {\em Phys. Rev. D} {\bf 1994}, {\em 50},~2225--2233. [\href{http://dx.doi.org/10.1103/PhysRevD.50.2225}{CrossRef}]

\bibitem[Ayala et~al.(1995)Ayala, Jalilian-Marian, McLerran, and
Venugopalan]{Ayala:1995kg}
Ayala, A.; Jalilian-Marian, J.; McLerran, L.D.; Venugopalan, R.
\newblock {The gluon propagator in non-Abelian Weizs\"acker--Williams fields}.
\newblock {\em Phys. Rev. D} {\bf 1995}, {\em 52},~2935--2943. [\href{http://dx.doi.org/10.1103/PhysRevD.52.2935}{CrossRef}]

\bibitem[Ayala et~al.(1996)Ayala, Jalilian-Marian, McLerran, and
Venugopalan]{Ayala:1995hx}
Ayala, A.; Jalilian-Marian, J.; McLerran, L.D.; Venugopalan, R.
\newblock {Quantum corrections to the Weizs\"acker--Williams gluon distribution
function at small $x$}.
\newblock {\em Phys. Rev. D} {\bf 1996}, {\em 53},~458--475. [\href{http://dx.doi.org/10.1103/PhysRevD.53.458}{CrossRef}]

\bibitem[Iancu and Venugopalan(2004)]{Iancu:2003xm}
Iancu, E.; Venugopalan, R.
\newblock The color glass condensate and high energy scattering in QCD. In {\em
Quark--Gluon Plasma 3}; {Hwa, R.C., Wang, X.-N., Eds.;}   World Scientific {Co. Ltd.}: {Singapore,} 
2004;  pp. 249--363. [\href{http://dx.doi.org/10.1142/9789812795533_0005}{CrossRef}]

\bibitem[Gelis et~al.(2010)Gelis, Iancu, Jalilian-Marian, and
Venugopalan]{Gelis:2010nm}
Gelis, F.; Iancu, E.; Jalilian-Marian, J.; Venugopalan, R.
\newblock {The Color Glass Condensate}.
\newblock {\em Ann. Rev. Nucl. Part. Sci.} {\bf 2010}, {\em 60},~463--489. [\href{http://dx.doi.org/10.1146/annurev.nucl.010909.083629}{CrossRef}]

\bibitem[Morreale and Salazar(2021)]{Morreale:2021pnn}
Morreale, A.; Salazar, F.
\newblock {Mining for Gluon Saturation at Colliders}.
\newblock {\em Universe} {\bf 2021}, {\em 7},~312. [\href{http://dx.doi.org/10.3390/universe7080312}{CrossRef}]

\bibitem[Nikolaev and Zakharov(1991)]{Nikolaev:1990ja}
Nikolaev, N.N.; Zakharov, B.G.
\newblock {Color transparency and scaling properties of nuclear shadowing in
deep inelastic scattering}.
\newblock {\em Z.~Phys. C} {\bf 1991}, {\em 49},~607--618. [\href{http://dx.doi.org/10.1007/BF01483577}{CrossRef}]

\bibitem[Nikolaev and Zakharov(1992)]{Nikolaev:1991et}
Nikolaev, N.{N.}; Zakharov, B.G.
\newblock {Pomeron structure function and diffraction dissociation of virtual
photons in perturbative QCD}.
\newblock {\em Z. Phys. C} {\bf 1992}, {\em 53},~331--346. [\href{http://dx.doi.org/10.1007/BF01597573}{CrossRef}]

\bibitem[Mueller(1994)]{Mueller:1993rr}
Mueller, A.H.
\newblock {Soft gluons in the infinite momentum wave function and the BFKL
pomeron}.
\newblock {\em Nucl. Phys. B} {\bf 1994}, {\em 415},~373--385. [\href{http://dx.doi.org/10.1016/0550-3213(94)90116-3}{CrossRef}]

\bibitem[Mueller and Patel(1994)]{Mueller:1994jq}
Mueller, A.H.; Patel, B.
\newblock {Single and double BFKL pomeron exchange and a dipole picture of
high-energy hard processes}.
\newblock {\em Nucl. Phys. B} {\bf 1994}, {\em 425},~471--488. [\href{http://dx.doi.org/10.1016/0550-3213(94)90284-4}{CrossRef}]

\bibitem[Henkels et~al.(2023)Henkels, de~Oliveira, Pasechnik, and
Trebien]{Henkels:2022bne}
Henkels, C.; de~Oliveira, E.G.; Pasechnik, R.; Trebien, H.
\newblock {Exclusive photo- and electroproduction of excited light vector
mesons via holographic model}.
\newblock {\em Eur. Phys. J. C} {\bf 2023}, {\em 83},~551. [\href{http://dx.doi.org/10.1140/epjc/s10052-023-11706-5}{CrossRef}]

\bibitem[Cisek et~al.(2023)Cisek, Sch\"afer, and Szczurek]{Cisek:2022yjj}
Cisek, A.; Sch\"afer, W.; Szczurek, A.
\newblock {Exclusive production of \ensuremath{\rho} meson in gamma-proton
collisions: d\ensuremath{\sigma}/dt and the role of helicity flip processes}.
\newblock {\em Phys. Lett. B} {\bf 2023}, {\em 836},~137595. [\href{http://dx.doi.org/10.1016/j.physletb.2022.137595}{CrossRef}]

\bibitem[M\"antysaari et~al.(2022)M\"antysaari, Salazar, and
Schenke]{Mantysaari:2022sux}
M\"antysaari, H.; Salazar, F.; Schenke, B.
\newblock {Nuclear geometry at high energy from exclusive vector meson
production}.
\newblock {\em Phys. Rev. D} {\bf 2022}, {\em 106},~074019. [\href{http://dx.doi.org/10.1103/PhysRevD.106.074019}{CrossRef}]

\bibitem[M\"antysaari and Penttala(2022{\natexlab{a}})]{Mantysaari:2022kdm}
M\"antysaari, H.; Penttala, J.
\newblock {Complete calculation of exclusive heavy vector meson production at
next-to-leading order in the dipole picture}.
\newblock {\em J. High Energy Phys.} {\bf 2022}, {\em 2022},~247. [\href{http://dx.doi.org/10.1007/JHEP08(2022)247}{CrossRef}]

\bibitem[M\"antysaari and Penttala(2022{\natexlab{b}})]{Mantysaari:2022bsp}
M\"antysaari, H.; Penttala, J.
\newblock {Exclusive production of light vector mesons at next-to-leading order
in the dipole picture}.
\newblock {\em Phys. Rev. D} {\bf 2022}, {\em 105},~114038. [\href{http://dx.doi.org/10.1103/PhysRevD.105.114038}{CrossRef}]

\bibitem[M\"antysaari and Penttala(2021)]{Mantysaari:2021ryb}
M\"antysaari, H.; Penttala, J.
\newblock {Exclusive heavy vector meson production at next-to-leading order in
the dipole picture}.
\newblock {\em Phys. Lett. B} {\bf 2021}, {\em 823},~136723. [\href{http://dx.doi.org/10.1016/j.physletb.2021.136723}{CrossRef}]

\bibitem[Kopeliovich et~al.(2021)Kopeliovich, Krelina, and
Nemchik]{Kopeliovich:2021dgx}
Kopeliovich, B.Z.; Krelina, M.; Nemchik, J.
\newblock {Electroproduction of heavy quarkonia: Significance of dipole
orientation}.
\newblock {\em Phys. Rev. D} {\bf 2021}, {\em 103},~094027. [\href{http://dx.doi.org/10.1103/PhysRevD.103.094027}{CrossRef}]

\bibitem[Kopeliovich et~al.(2022)Kopeliovich, Krelina, Nemchik, and
Potashnikova]{Kopeliovich:2022jwe}
Kopeliovich, B.Z.; Krelina, M.; Nemchik, J.; Potashnikova, I.K.
\newblock {Coherent photoproduction of heavy quarkonia on nuclei}.
\newblock {\em Phys. Rev. D} {\bf 2022}, {\em 105},~054023. [\href{http://dx.doi.org/10.1103/PhysRevD.105.054023}{CrossRef}]

\bibitem[Cepila et~al.(2023)Cepila, Contreras, and Vaculciak]{Cepila:2023pvh}
Cepila, J.; Contreras, J.G.; Vaculciak, M.
\newblock {Solutions to the Balitsky--Kovchegov equation including the dipole
orientation}. \emph{Phys. Lett. B} {\bf 2024}, \emph{848}, {138360}. [\href{http://dx.doi.org/10.1016/j.physletb.2023.138360}{CrossRef}]

\bibitem[Matousek et~al.(2023)Matousek, Khachatryan, and
Zhang]{Matousek:2022enl}
Matousek, G.; Khachatryan, V.; Zhang, J.
\newblock {Scaling properties of exclusive vector meson production cross
section from gluon saturation}.
\newblock {\em Eur. Phys. J. Plus} {\bf 2023}, {\em 138},~113. [\href{http://dx.doi.org/10.1140/epjp/s13360-023-03729-4}{CrossRef}]

\bibitem[Kumar and Toll(2022)]{Kumar:2022aly}
Kumar, A.; Toll, T.
\newblock {Energy dependence of the proton geometry in exclusive vector meson
production}.
\newblock {\em Phys. Rev. D} {\bf 2022}, {\em 105},~114011. [\href{http://dx.doi.org/10.1103/PhysRevD.105.114011}{CrossRef}]

\bibitem[Anand and Toll(2019)]{Anand:2018zle}
Anand, S.; Toll, T.
\newblock {Exclusive diffractive vector meson production: A comparison between
the dipole model and the leading twist shadowing approach}.
\newblock {\em Phys. Rev. C} {\bf 2019}, {\em 100},~024901. [\href{http://dx.doi.org/10.1103/PhysRevC.100.024901}{CrossRef}]

\bibitem[Xie and Goncalves(2022)]{Xie:2022sjm}
Xie, Y.P.; Goncalves, V.P.
\newblock {Exclusive processes in ep collisions at the EIC and LHeC: A closer
look on the predictions of saturation models}.
\newblock {\em Phys. Rev. D} {\bf 2022}, {\em 105},~014033. [\href{http://dx.doi.org/10.1103/PhysRevD.105.014033}{CrossRef}]

\bibitem[Henkels et~al.(2023)Henkels, de~Oliveira, Pasechnik, and
Trebien]{Henkels:2023plt}
Henkels, C.; de~Oliveira, E.G.; Pasechnik, R.; Trebien, H.
\newblock {Coherent photoproduction of light vector mesons off nuclear targets
in the dipole picture}. \emph{Nucl. Phys. A} {\bf 2025}, \emph{1055}, {123018}. [\href{http://dx.doi.org/10.1016/j.nuclphysa.2025.123018}{CrossRef}]

\bibitem[M\"antysaari et~al.(2023{\natexlab{a}})M\"antysaari, Salazar, and
Schenke]{Mantysaari:2023xcu}
M\"antysaari, H.; Salazar, F.; Schenke, B.
\newblock {Energy dependent nuclear suppression from gluon saturation in
exclusive vector meson production}. \emph{Phys. Rev. D} {\bf 2024},   \emph{109}, {L071504}. [\href{http://dx.doi.org/10.1103/PhysRevD.109.L071504}{CrossRef}]

\bibitem[M\"antysaari et~al.(2023{\natexlab{b}})M\"antysaari, Salazar, Schenke,
Shen, and Zhao]{Mantysaari:2023prg}
M\"antysaari, H.; Salazar, F.; Schenke, B.; Shen, C.; Zhao, W.
\newblock {Effects of nuclear structure and quantum interference on diffractive
vector meson production in ultraperipheral nuclear collisions}. \emph{Phys. Rev. C} {\bf 2024}, \emph{109}, {024908}. [\href{http://dx.doi.org/10.1103/PhysRevC.109.024908}{CrossRef}]

\bibitem[Azevedo et~al.(2023)Azevedo, Goncalves, and Moreira]{Azevedo:2023fee}
Azevedo, C.N.; Goncalves, V.P.; Moreira, B.D.
\newblock {Associated vector-meson and bound-free electron-positron pair
photoproduction in ultraperipheral PbPb collisions}.
\newblock {\em Phys. Rev. C} {\bf 2023}, {\em 108},~064907. [\href{http://dx.doi.org/10.1103/PhysRevC.108.064907}{CrossRef}]

\bibitem[Goncalves et~al.(2023)Goncalves, Klasen, and
Moreira]{Goncalves:2023sts}
Goncalves, V.P.; Klasen, M.; Moreira, B.D.
\newblock {Exclusive $J/\Psi$ plus jet associated production in ultraperipheral
$PbPb$ collisions}.
\newblock {\em Eur. Phys. J. C} {\bf 2023}, {\em 83},~895. [\href{http://dx.doi.org/10.1140/epjc/s10052-023-12068-8}{CrossRef}]

\bibitem[Boroun(2024)]{Boroun:2023vuu}
Boroun, G.R.
\newblock {A simple model for the charm structure function of nuclei}.
\newblock {\em Phys. Lett. B} {\bf 2024}, {\em 849},~138440. [\href{http://dx.doi.org/10.1016/j.physletb.2023.138440}{CrossRef}]

\bibitem[Boroun(2023)]{Boroun:2023klw}
Boroun, G.R.
\newblock {Heavy quark structure functions from unifying the color dipole
picture and double asymptotic scaling approaches}. \emph{Phys. Rev. D} {\bf 2024}, \emph{109}, {054012}. [\href{http://dx.doi.org/10.1103/PhysRevD.109.054012}{CrossRef}]


\bibitem[Bertulani and Baur(1988)]{Bertulani:1987tz}
Bertulani, C.A.; Baur, G.
\newblock {Electromagnetic processes in relativistic heavy ion collisions}.
\newblock {\em Phys. Rep.} {\bf 1988}, {\em 163},~299{--408}. [\href{http://dx.doi.org/10.1016/0370-1573(88)90142-1}{CrossRef}]

\bibitem[Baur et~al.(2002)Baur, Hencken, Trautmann, Sadovsky, and
Kharlov]{Baur:2001jj}
Baur, G.; Hencken, K.; Trautmann, D.; Sadovsky, S.; Kharlov, Y.
\newblock {Coherent {$\gamma \gamma$ and $\gamma A$} interactions in very peripheral
collisions at relativistic ion colliders}.
\newblock {\em Phys. Rep.} {\bf 2002}, {\em 364},~359--450. [\href{http://dx.doi.org/10.1016/S0370-1573(01)00101-6}{CrossRef}]

\bibitem[Bertulani et~al.(2005)Bertulani, Klein, and
Nystrand]{Bertulani:2005ru}
Bertulani, C.A.; Klein, S.R.; Nystrand, J.
\newblock {Physics of ultra-peripheral nuclear collisions}.
\newblock {\em Ann. Rev. Nucl. Part. Sci.} {\bf 2005}, {\em 55},~271--310. [\href{http://dx.doi.org/10.1146/annurev.nucl.55.090704.151526}{CrossRef}]

\bibitem[Baltz(2008)]{Baltz:2007kq}
Baltz, A.J.; {Baur, G.; d'Enterria, D.; Frankfurt, L.; Gelis, F.; Guzey, V.; Hencken, K.; Kharlov, Y.; Klasen, M.; Klein, S.R.; et al.} 
\newblock {The physics of ultraperipheral collisions at the LHC}.
\newblock {\em Phys. Rep.} {\bf 2008}, {\em 458},~1--171. [\href{http://dx.doi.org/10.1016/j.physrep.2007.12.001}{CrossRef}]

\bibitem[Klein and Steinberg(2020)]{Klein:2020fmr}
Klein, S.; Steinberg, P.
\newblock {Photonuclear and two-photon interactions at high-energy nuclear
colliders}.
\newblock {\em Ann. Rev. Nucl. Part. Sci.} {\bf 2020}, {\em 70},~323--354. [\href{http://dx.doi.org/10.1146/annurev-nucl-030320-033923}{CrossRef}]

\bibitem[Fagundes and Machado(2023)]{Fagundes:2022bzw}
Fagundes, D.A.; Machado, M.V.T.
\newblock {Asymptotic gluon density within the color dipole picture in light of
HERA high-precision data}.
\newblock {\em Phys. Rev. D} {\bf 2023}, {\em 107},~014004. [\href{http://dx.doi.org/10.1103/PhysRevD.107.014004}{CrossRef}]

\bibitem[Caola and Forte(2008)]{Caola:2008xr}
Caola, F.; Forte, S.
\newblock {Geometric scaling from GLAP evolution}.
\newblock {\em Phys. Rev. Lett.} {\bf 2008}, {\em 101},~022001. [\href{http://dx.doi.org/10.1103/PhysRevLett.101.022001}{CrossRef}]

\bibitem[Ball et~al.(2016)Ball, Nocera, and Rojo]{Ball:2016spl}
Ball, R.D.; Nocera, E.R.; Rojo, J.
\newblock {The asymptotic behaviour of parton distributions at small and large
$x$}.
\newblock {\em Eur. Phys. J. C} {\bf 2016}, {\em 76},~383. [\href{http://dx.doi.org/10.1140/epjc/s10052-016-4240-4}{CrossRef}]

\bibitem[Rezaeian and Schmidt(2013)]{Rezaeian:2013tka}
Rezaeian, A.H.; Schmidt, I.
\newblock {Impact-parameter dependent color glass condensate dipole model and
new combined HERA data}.
\newblock {\em Phys. Rev. D} {\bf 2013}, {\em 88},~074016. [\href{http://dx.doi.org/10.1103/PhysRevD.88.074016}{CrossRef}]

\bibitem[Mäntysaari and Zurita(2018)]{Mantysaari:2018nng}
M\"antysaari, H.; Zurita, P.
\newblock {In depth analysis of the combined HERA data in the dipole models
with and without saturation}.
\newblock {\em Phys. Rev. D} {\bf 2018}, {\em 98},~036002. [\href{http://dx.doi.org/10.1103/PhysRevD.98.036002}{CrossRef}]

\bibitem[Ball and Forte(1994)]{Ball:1994du}
Ball, R.D.; Forte, S.
\newblock {Double asymptotic scaling at HERA}.
\newblock {\em Phys. Lett. B} {\bf 1994}, {\em 335},~77--86. [\href{http://dx.doi.org/10.1016/0370-2693(94)91561-X}{CrossRef}]

\bibitem[Kowalski and Teaney(2003)]{Kowalski:2003hm}
Kowalski, H.; Teaney, D.
\newblock {An impact parameter dipole saturation model}.
\newblock {\em Phys. Rev. D} {\bf 2003}, {\em 68},~114005. [\href{http://dx.doi.org/10.1103/PhysRevD.68.114005}{CrossRef}]

\bibitem[Mäntysaari and Schenke(2018)]{Mantysaari:2018zdd}
M\"antysaari, H.; Schenke, B.
\newblock {Confronting impact parameter dependent JIMWLK evolution with HERA
data}.
\newblock {\em Phys. Rev. D} {\bf 2018}, {\em 98},~034013. [\href{http://dx.doi.org/10.1103/PhysRevD.98.034013}{CrossRef}]

\bibitem[Golec-Biernat and Wusthoff(1998)]{Golec-Biernat:1998zce}
Golec-Biernat, K.J.; Wusthoff, M.
\newblock {Saturation effects in deep inelastic scattering at low $Q^2$ and its
implications on diffraction}.
\newblock {\em Phys. Rev. D} {\bf 1998}, {\em 59},~014017. [\href{http://dx.doi.org/10.1103/PhysRevD.59.014017}{CrossRef}]

\bibitem[Golec-Biernat and Wusthoff(1999)]{GolecBiernat:1999qd}
Golec-Biernat, K.J.; Wusthoff, M.
\newblock {Saturation in diffractive deep inelastic scattering}.
\newblock {\em Phys. Rev. D} {\bf 1999}, {\em 60},~114023. [\href{http://dx.doi.org/10.1103/PhysRevD.60.114023}{CrossRef}]

\bibitem[Bartels et~al.(2002)Bartels, Golec-Biernat, and
Kowalski]{Bartels:2002cj}
Bartels, J.; Golec-Biernat, K.J.; Kowalski, H.
\newblock {A modification of the saturation model: DGLAP evolution}.
\newblock {\em Phys. Rev. D} {\bf 2002}, {\em 66},~014001. [\href{http://dx.doi.org/10.1103/PhysRevD.66.014001}{CrossRef}]

\bibitem[Golec-Biernat and Sapeta(2006)]{Golec-Biernat:2006koa}
Golec-Biernat, K.J.; Sapeta, S.
\newblock {Heavy flavour production in DGLAP improved saturation model}.
\newblock {\em Phys. Rev. D} {\bf 2006}, {\em 74},~054032. [\href{http://dx.doi.org/10.1103/PhysRevD.74.054032}{CrossRef}]

\bibitem[Golec-Biernat and Sapeta(2018)]{Golec-Biernat:2017lfv}
Golec-Biernat, K.; Sapeta, S.
\newblock {Saturation model of DIS: An update}.
\newblock {\em J. High Energy Phys.} {\bf 2018}, {\em 2018},~102. [\href{http://dx.doi.org/10.1007/JHEP03(2018)102}{CrossRef}]

\bibitem[\L{}uszczak et~al.(2022)\L{}uszczak, \L{}uszczak, and
Sch\"afer]{Luszczak:2022fkf}
\L{}uszczak, A.; \L{}uszczak, M.; Sch\"afer, W.
\newblock {Unintegrated gluon distributions from the color dipole cross section
in the BGK saturation model}.
\newblock {\em Phys. Lett. B} {\bf 2022}, {\em 835},~137582. [\href{http://dx.doi.org/10.1016/j.physletb.2022.137582}{CrossRef}]

\bibitem[Boroun(2023)]{Boroun:2023ldq}
Boroun, G.R.
\newblock {The unintegrated gluon distribution from the GBW and BGK models}. \emph{Eur. Phys. J. A} {\bf 2024}, \emph{60}, {48}. [\href{http://dx.doi.org/10.1140/epja/s10050-024-01255-0}{CrossRef}]

\bibitem[Kovchegov and Levin(2013)]{Kovchegov:2012mbw}
Kovchegov, Y.V.; Levin, E.
\newblock {\em Quantum Chromodynamics at High Energy}; Cambridge University Press: {Cambridge, UK,}
2013. [\href{http://dx.doi.org/10.1017/CBO9781139022187}{CrossRef}]

\bibitem[Kowalski et~al.(2006)Kowalski, Motyka, and Watt]{Kowalski:2006hc}
Kowalski, H.; Motyka, L.; Watt, G.
\newblock {Exclusive diffractive processes at HERA within the dipole picture}.
\newblock {\em Phys. Rev. D} {\bf 2006}, {\em 74},~074016. [\href{http://dx.doi.org/10.1103/PhysRevD.74.074016}{CrossRef}]

\bibitem[Shuvaev et~al.(1999)Shuvaev, Golec-Biernat, Martin, and
Ryskin]{Shuvaev:1999ce}
Shuvaev, A.G.; Golec-Biernat, K.J.; Martin, A.D.; Ryskin, M.G.
\newblock {Off diagonal distributions fixed by diagonal partons at small $x$ and
$\xi$}.
\newblock {\em Phys. Rev. D} {\bf 1999}, {\em 60},~014015. [\href{http://dx.doi.org/10.1103/PhysRevD.60.014015}{CrossRef}]

\bibitem[Harland-Lang(2013)]{Harland-Lang:2013xba}
Harland-Lang, L.A.
\newblock {Simple form for the low-$x$ generalized parton distributions in the
skewed regime}.
\newblock {\em Phys. Rev. D} {\bf 2013}, {\em 88},~034029. [\href{http://dx.doi.org/10.1103/PhysRevD.88.034029}{CrossRef}]

%\cite{Favart:2005sc}
\bibitem{Favart:2005sc}
Favart, L.; Machado, M.V.T.; Schoeffel, L.
Extraction of the skewing factor from the DIS/DVCS ratio. \emph{arXiv} \textbf{2005}, arXiv:hep-ph/0511069. [\href{http://dx.doi.org/10.48550/arXiv.hep-ph/0511069}{CrossRef}]
%6 citations counted in INSPIRE as of 19 May 2025

%\cite{Favart:2007zz}
\bibitem{Favart:2007zz}
Favart, L.; Machado, M.V.T.; Schoeffel, L.
{An} 
extraction of the skewing factor from DESY-HERA data.
\emph{Braz. J. Phys.} \textbf{2007}, \emph{37}, 798--800. [\href{http://dx.doi.org/10.1590/S0103-97332007000500031}{CrossRef}]
%4 citations counted in INSPIRE as of 19 May 2025

%\cite{Schoeffel:2008xw}
\bibitem{Schoeffel:2008xw}
Schoeffel, L.
{Deeply} virtual compton scattering at HERA and perspectives at CERN.
\emph{AIP Conf. Proc.} \textbf{2008}, \emph{1056}, 372--379. [\href{http://dx.doi.org/10.1063/1.3013066}{CrossRef}]
%[arXiv:0806.3047 [hep-ph]].
%0 citations counted in INSPIRE as of 19 May 2025

\bibitem[Nemchik et~al.(1994)Nemchik, Nikolaev, and Zakharov]{Nemchik:1994fp}
Nemchik, J.; Nikolaev, N.N.; Zakharov, B.G.
\newblock {Scanning the BFKL pomeron in elastic production of vector mesons at
HERA}.
\newblock {\em Phys. Lett. B} {\bf 1994}, {\em 341},~228--237. [\href{http://dx.doi.org/10.1016/0370-2693(94)90314-X}{CrossRef}]

\bibitem[Nemchik et~al.(1997)Nemchik, Nikolaev, Predazzi, and
Zakharov]{Nemchik:1996cw}
Nemchik, J.; Nikolaev, N.N.; Predazzi, E.; Zakharov, B.G.
\newblock {Color dipole phenomenology of diffractive electroproduction of light
vector mesons at HERA}.
\newblock {\em Z. Phys. C} {\bf 1997}, {\em 75},~71--87. [\href{http://dx.doi.org/10.1007/s002880050448}{CrossRef}]

\bibitem[Cox et~al.(2009)Cox, Forshaw, and Sandapen]{Cox:2009ag}
Cox, B.E.; Forshaw, J.R.; Sandapen, R.
\newblock {Diffractive {$\Upsilon$ 
production at the Tevatron and} LHC}.
\newblock {\em J. High Energy Phys.} {\bf 2009}, {\em 2009},~
{34}. [\href{http://dx.doi.org/10.1088/1126-6708/2009/06/034}{CrossRef}]

\bibitem[Armesto and Rezaeian(2014)]{Armesto:2014sma}
Armesto, N.; Rezaeian, A.H.
\newblock {Exclusive vector meson production at high energies and gluon
saturation}.
\newblock {\em Phys. Rev. D} {\bf 2014}, {\em 90},~054003. [\href{http://dx.doi.org/10.1103/PhysRevD.90.054003}{CrossRef}]

%\cite{SampaiodosSantos:2014puz}
\bibitem{SampaiodosSantos:2014puz}
{Sampaio} dos Santos, G.
; Machado, M.V.T.
{On} theoretical uncertainty of color dipole phenomenology in the $J/\psi $ and $\Upsilon$ photoproduction in {$pA$ and $AA$} collisions at the CERN
Large
Hadron Collider.
\emph{J. Phys. G} \textbf{2015}, \emph{42}, 105001. [\href{http://dx.doi.org/10.1088/0954-3899/42/10/105001}{CrossRef}]
%[arXiv:1411.7918 [hep-ph]].
%50 citations counted in INSPIRE as of 19 May 2025

%\cite{SampaiodosSantos:2014qtt}
\bibitem{SampaiodosSantos:2014qtt}
{Sampaio} dos Santos, G.
; Machado, M.V.T.
{Light} vector meson photoproduction in hadron--hadron and nucleus--nucleus
collisions at energies available at the CERN Large Hadron Collider.
\emph{Phys. Rev. C} \textbf{2015}, \emph{91}, 025203. [\href{http://dx.doi.org/10.1103/PhysRevC.91.025203}{CrossRef}]
%[arXiv:1407.4148 [hep-ph]].
%38 citations counted in INSPIRE as of 19 May 2025

%\cite{Goncalves:2017wgg}
\bibitem{Goncalves:2017wgg}
Gon\c{c}alves, V.P.; Machado, M.V.T.; Moreira, B.D.; Navarra, F.S.; {Sampaio} dos Santos, G. {Color} dipole predictions for the exclusive vector meson photoproduction in $pp$, $p$Pb, and PbPb collisions at run 2 LHC energies.
\emph{Phys. Rev. D} \textbf{2017}, \linebreak  \emph{96}, 094027. [\href{http://dx.doi.org/10.1103/PhysRevD.96.094027}{CrossRef}]
%[arXiv:1710.10070 [hep-ph]].
%82 citations counted in INSPIRE as of 19 May 2025



\bibitem[Frankfurt et~al.(1998)Frankfurt, Guzey, and
Strikman]{Frankfurt:1997zk}
Frankfurt, L.; Guzey, V.; Strikman, M.
\newblock {{Cross section} fluctuations of photon projectile in {the} generalized
vector meson dominance model}.
\newblock {\em Phys. Rev. D} {\bf 1998}, {\em 58},~094039. [\href{http://dx.doi.org/10.1103/PhysRevD.58.094039}{CrossRef}]

\bibitem[Forshaw et~al.(1999)Forshaw, Kerley, and Shaw]{Forshaw:1999uf}
Forshaw, J.R.; Kerley, G.; Shaw, G.
\newblock {Extracting the dipole {cross section from photo- 
and} electroproduction total cross-section data}.
\newblock {\em Phys. Rev. D} {\bf 1999}, {\em 60},~074012. [\href{http://dx.doi.org/10.1103/PhysRevD.60.074012}{CrossRef}]

\bibitem[Gon\c{c}alves and Moreira(2020)]{Goncalves:2020cir}
Gon\c{c}alves, V.P.; Moreira, B.D.
\newblock {A phenomenological analysis of the nonperturbative QCD contributions
for the photon wave function}.
\newblock {\em Eur. Phys. J. C} {\bf 2020}, {\em 80},~492. [\href{http://dx.doi.org/10.1140/epjc/s10052-020-8043-2}{CrossRef}]

\bibitem[Glauber et~al.(1959)]{glauber1959lectures}
Glauber, R.{J.}  
{High-energy collision theory.}
\newblock {In \em {\it Lectures in Theoretical Physics. {Volume 1}}}; {Brittin, W.E., Dunham, L.G., Eds.}; Interscience {Publishers, Inc.}:
New York, NY, USA; London, UK, 1959{; pp. 315--414}.
{Available online:} \url{https://archive.org/details/lecturesintheore0001unse/} (accessed on 31 March 2025).

\bibitem[Gribov(1969{\natexlab{a}})]{Gribov:1968jf}
Gribov, V.N.
\newblock {Glauber corrections and the interaction between high-energy hadrons
and nuclei}.
\newblock {\em Sov. Phys. JETP} {\bf 1969}, {\em 29},~483--487.
{Available online:} \url{http://jetp.ras.ru/cgi-bin/e/index/e/29/3/p483?a=list} (accessed on 31 March 2025).


\bibitem[Gribov(1969{\natexlab{b}})]{Gribov:1968gs}
Gribov, V.N.
\newblock {Interaction of gamma quanta and electrons with nuclei at high energies}.
\newblock
{\em Sov. Phys. JETP.} {\bf 1970}, {\em 30},~{709--717.} 
%{\em Zh. Eksp. Teor. Fiz.} {\bf 1969}, {\em 57},~1306--1323.
{Available online:} \url{http://jetp.ras.ru/cgi-bin/e/index/e/30/4/p709?a=list} (accessed on 31 March 2025).

\bibitem[Armesto(2002)]{Armesto:2002ny}
Armesto, N.
\newblock {A Simple model for nuclear structure functions at small x in the
dipole picture}.
\newblock {\em Eur. Phys. J. C} {\bf 2002}, {\em 26},~35--43. [\href{http://dx.doi.org/10.1007/s10052-002-1021-z}{CrossRef}]

\bibitem[Woods and Saxon(1954)]{PhysRev.95.577}
Woods, R.D.; Saxon, D.S.
\newblock Diffuse surface optical model for nucleon--nuclei scattering.
\newblock {\em Phys. Rev.} {\bf 1954}, {\em 95},~577--578. [\href{http://dx.doi.org/10.1103/PhysRev.95.577}{CrossRef}]
\clearpage
\bibitem[{De Vries} et~al.(1987){De Vries}, {De Jager}, and {De
Vries}]{DEVRIES1987495}
{De Vries}, H.; {De Jager}, C.; {De Vries}, C.
\newblock Nuclear charge-density-distribution parameters from elastic electron
scattering.
\newblock {\em Atom. Data Nucl. Data Tables} {\bf 1987}, {\em
36},~495--536. [\href{http://dx.doi.org/10.1016/0092-640X(87)90013-1}{CrossRef}]

%\cite{Roa:2023skv}
\bibitem{Roa:2023skv}
Roa, M.; Garrido, J.; Guevara, M.
{CGC} and saturation approach: Impact-parameter dependent model of perturbative QCD and combined HERA data.
\emph{Phys. Rev. D} \textbf{2024}, \emph{110}, 074006. [\href{http://dx.doi.org/10.1103/PhysRevD.110.074006}{CrossRef}]
%[arXiv:2311.06406 [hep-ph]].
%2 citations counted in INSPIRE as of 19 May 2025


%\cite{Drees:1988pp}
\bibitem{Drees:1988pp}
Drees, M.; Zeppenfeld, D.
{Production} of supersymmetric particles in elastic $ep$ collisions.
\emph{Phys. Rev. D} \textbf{1989}, \emph{39}, 2536. [\href{http://dx.doi.org/10.1103/PhysRevD.39.2536}{CrossRef}]
%181 citations counted in INSPIRE as of 19 May 2025

\bibitem[Aaij et~al.(2018)]{LHCb:2018rcm}
Aaij, R. et~al. {[The LHCb Collaboration]} 
\newblock {Central exclusive production of $J/\psi$ and $\psi(2S)$ mesons in
$pp$ collisions at $\sqrt{s}=13~$TeV}.
\newblock {\em J. High Energy Phys.} {\bf 2018}, {\em 2018
},~167. [\href{http://dx.doi.org/10.1007/JHEP10(2018)167}{CrossRef}]

\bibitem[Jones et~al.(2017)Jones, Martin, Ryskin, and Teubner]{Jones:2016icr}
Jones, S.P.; Martin, A.D.; Ryskin, M.G.; Teubner, T.
\newblock {Exclusive $J/\psi$ production at the LHC in the $k_T$ factorization
approach}.
\newblock {\em J. Phys. G} {\bf 2017}, {\em 44},~03LT01. [\href{http://dx.doi.org/10.1088/1361-6471/aa56ea}{CrossRef}]

\bibitem[Abelev et~al.(2014)]{ALICE:2014eof}
Abelev, B.B. et~al. {[ALICE Collaboration]}
\newblock {Exclusive $J/\psi$ photoproduction off protons in
ultraperipheral $p$--Pb collisions at $\sqrt{s_{\rm NN}}=5.02$ TeV}.
\newblock {\em Phys. Rev. Lett.} {\bf 2014}, {\em 113},~232504. [\href{http://dx.doi.org/10.1103/PhysRevLett.113.232504}{CrossRef}]

\bibitem[Acharya et~al.(2019)]{ALICE:2018oyo}
Acharya, S. et~al. [{ALICE Collaboration}]
\newblock {Energy dependence of exclusive $\mathrm {J}/\psi $ photoproduction
off protons in ultra-peripheral p--Pb collisions at
$\sqrt{s_{\mathrm{NN}}} = 5.02$ TeV}.
\newblock {\em Eur. Phys. J. C} {\bf 2019}, {\em 79},~402. [\href{http://dx.doi.org/10.1140/epjc/s10052-019-6816-2}{CrossRef}]

\bibitem[Sirunyan et~al.(2019)]{CMS:2018bbk}
Sirunyan, A.M. et~al. {[CMS Collaboration]}
\newblock {Measurement of exclusive $\Upsilon$ photoproduction from protons in
pPb collisions at $\sqrt{s_\mathrm{NN}} =$ 5.02 TeV}.
\newblock {\em Eur. Phys. J. C} {\bf 2019}, {\em 79},~{277.} 
%\newblock Erratum in: \emph{Eur. Phys. J. C} \textbf{2022}, \emph{82}, 343. [\href{http://dx.doi.org/10.1140/epjc/s10052-019-6774-8}{CrossRef}]

\bibitem[Flett et~al.(2022)Flett, Jones, Martin, Ryskin, and
Teubner]{Flett:2022ues}
Flett, C.A.; Jones, S.P.; Martin, A.D.; Ryskin, M.G.; Teubner, T.
\newblock {Exclusive $J/\psi$ and $\Upsilon$ production in high energy $pp$ and
$p$--Pb collisions}.
\newblock {\em Phys. Rev. D} {\bf 2022}, {\em 106},~074021. [\href{http://dx.doi.org/10.1103/PhysRevD.106.074021}{CrossRef}]

\bibitem[Tumasyan et~al.(2023)]{CMS:2023snh}
Tumasyan, A. et~al. {[CMS Collaboration]}
\newblock {Probing small Bjorken-$x$ nuclear gluonic structure via coherent
$J/\psi$ photoproduction in ultraperipheral Pb--Pb collisions at
$\sqrt{s_\mathrm{NN}}$ = 5.02 TeV}.
%\emph{arXiv} {\bf 2023}, arXiv:2303.16984.
\emph{Phys. Rev. Lett.} {\bf 2023}, {262301.} 
. [\href{http://dx.doi.org/10.1103/PhysRevLett.131.262301}{CrossRef}]


%
% NEW REFERENCES FOR FIG. 3.
%

\bibitem{H1:2000kis}
Adloff, C. {et al.} [H1 Collaboration] 
Elastic photoproduction of $J/ \psi$ and $\Upsilon$ mesons at HERA. 
{\it Phys. Lett. B} \textbf{2000}, {\it 483}, 23--35.  [\href{http://dx.doi.org/10.1016/S0370-2693(00)00530-X}{CrossRef}]
%[arXiv:hep-ex/0003020 [hep-ex]].


\bibitem{ZEUS:2009asc}
Chekanov, S. {et al.} [ZEUS Collaboration] 
Exclusive photoproduction of $\Upsilon$ mesons at HERA. 
{\it Phys. Lett. B} \textbf{2009}, {\it 680}, 4--12. [\href{http://dx.doi.org/10.1016/j.physletb.2009.07.066}{CrossRef}]
%[arXiv:0903.4205 [hep-ex]].

\bibitem{ZEUS:1998cdr}
Breitweg, J. {et al.} [ZEUS Collaboration] 
Measurement of elastic $\Upsilon$ photoproduction at HERA. 
{\it Phys. Lett. B} \textbf{1998}, {\it 437}, 432--444. [\href{http://dx.doi.org/10.1016/S0370-2693(98)01081-8}{CrossRef}]
%[arXiv:hep-ex/9807020 [hep-ex]].

%

\bibitem[Acharya et~al.(2019)]{ALICE:2019tqa}
Acharya, S. et~al. {[ALICE Collaboration]}
\newblock {Coherent J/$\psi$ photoproduction at forward rapidity in
ultra-peripheral Pb--Pb collisions at $\sqrt{s_{\rm{NN}}}=5.02$ TeV}.
\newblock {\em Phys. Lett. B} {\bf 2019}, {\em 798},~134926. [\href{http://dx.doi.org/10.1016/j.physletb.2019.134926}{CrossRef}]

\bibitem[Acharya et~al.(2021)]{ALICE:2021gpt}
Acharya, S. et~al. {[ALICE Collaboration]}
\newblock {Coherent J/$\psi$ and $\psi'$ photoproduction at midrapidity in
ultra-peripheral Pb--Pb collisions at $\sqrt{s_{\mathrm{NN}}}~=~5.02$ TeV}.
\newblock {\em Eur. Phys. J. C} {\bf 2021}, {\em 81},~712. [\href{http://dx.doi.org/10.1140/epjc/s10052-021-09437-6}{CrossRef}]

\bibitem[Bursche(2019)]{Bursche:2018eni}
Bursche, A. {(on behalf of the LHCb Collaboration)}.
\newblock {Study of coherent $J/\psi$ production in lead--lead collisions at
$\sqrt{s_{\rm NN}} =5\ \rm{TeV}$ with the LHCb experiment}.
\newblock {\em Nucl. Phys. A} {\bf 2019}, {\em 982},~247--250. [\href{http://dx.doi.org/10.1016/j.nuclphysa.2018.10.069}{CrossRef}]

\bibitem[\L{}uszczak and Sch\"afer(2022)]{Luszczak:2021jtr}
\L{}uszczak, A.; Sch\"afer, W.
\newblock {Coherent photoproduction of $J/\psi$ in nucleus-nucleus collisions
in the color dipole approach---An update}.
\newblock {\em SciPost Phys. Proc.} {\bf 2022}, {\em 8},~109. [\href{http://dx.doi.org/10.21468/SciPostPhysProc.8.109}{CrossRef}]

\bibitem[Henkels et~al.(2021)Henkels, de~Oliveira, Pasechnik, and
Trebien]{Henkels:2020qvo}
Henkels, C.; de~Oliveira, E.G.; Pasechnik, R.; Trebien, H.
\newblock {Momentum transfer squared dependence of exclusive quarkonia
photoproduction in ultraperipheral collisions}.
\newblock {\em Phys. Rev. D} {\bf 2021}, {\em 104},~054008. [\href{http://dx.doi.org/10.1103/PhysRevD.104.054008}{CrossRef}]

\bibitem[Henkels et~al.(2020)Henkels, de~Oliveira, Pasechnik, and
Trebien]{Henkels:2020kju}
Henkels, C.; de~Oliveira, E.G.; Pasechnik, R.; Trebien, H.
\newblock {Exclusive photoproduction of excited quarkonia in ultraperipheral
collisions}.
\newblock {\em Phys. Rev. D} {\bf 2020}, {\em 102},~014024. [\href{http://dx.doi.org/10.1103/PhysRevD.102.014024}{CrossRef}]

\bibitem[Ducati et~al.(2013)Ducati, Griep, and Machado]{Ducati:2013bya}
Ducati, M.B.G.; Griep, M.T.; Machado, M.V.T.
\newblock {Diffractive photoproduction of radially excited {$\psi(2S)$} mesons in
photon--Pomeron reactions in PbPb collisions at the CERN {Large Hadron Collider}}.
\newblock {\em Phys. Rev. C} {\bf 2013}, {\em 88},~014910. [\href{http://dx.doi.org/10.1103/PhysRevC.88.014910}{CrossRef}]

\bibitem[Kopeliovich et~al.(2000)Kopeliovich, Schafer, and
Tarasov]{Kopeliovich:1999am}
{Kopeliovich, B.
;
Sch\"afer, A.; Tarasov, A.
} 
\newblock {Nonperturbative effects in gluon radiation and photoproduction of
quark pairs}.
\newblock {\em Phys. Rev. D} {\bf 2000}, {\em 62},~054022. [\href{http://dx.doi.org/10.1103/PhysRevD.62.054022}{CrossRef}]

\bibitem[Kopeliovich et~al.(2023)Kopeliovich, Krelina, Nemchik, and
Potashnikova]{Kopeliovich:2020has}
Kopeliovich, B.Z.; Krelina, M.; Nemchik, J.; Potashnikova, I.K.
\newblock {Ultraperipheral nuclear collisions as a source of heavy quarkonia}.
\newblock {\em Phys. Rev. D} {\bf 2023}, {\em 107},~054005. [\href{http://dx.doi.org/10.1103/PhysRevD.107.054005}{CrossRef}]

\bibitem[Acharya et~al.(2020)]{ALICE:2020ugp}
Acharya, S. et~al. {[ALICE collaboration]}
\newblock {Coherent photoproduction of $\rho^{0}$ vector mesons in
ultra-peripheral Pb--Pb collisions at $ \sqrt{{s}_{\mathrm{NN}}} $ =
5.02 TeV}.
\newblock {\em J. High Energy Phys.} {\bf 2020}, {\em 2020},~
{35}. [\href{http://dx.doi.org/10.1007/JHEP06(2020)035}{CrossRef}]

\bibitem[Acharya et~al.(2021)]{ALICE:2021jnv}
Acharya, S. et~al. {[ALICE Collaboration]}
\newblock {First measurement of coherent $\rho^0$ photoproduction in
ultra-peripheral Xe--Xe collisions at $ \sqrt{{s}_{\mathrm{NN}}} $ = 5.44 TeV}.
\newblock {\em Phys. Lett. B} {\bf 2021}, {\em 820},~136481. [\href{http://dx.doi.org/10.1016/j.physletb.2021.136481}{CrossRef}]

\bibitem[Baltz et~al.(2002)Baltz, Klein, and Nystrand]{Baltz:2002pp}
Baltz, A.J.; Klein, S.R.; Nystrand, J.
\newblock {Coherent {vector--meson} photoproduction with nuclear breakup in
relativistic {heavy--ion} collisions}.
\newblock {\em Phys. Rev. Lett.} {\bf 2002}, {\em 89},~012301. [\href{http://dx.doi.org/10.1103/PhysRevLett.89.012301}{CrossRef}]

\bibitem[Broz et~al.(2020)Broz, Contreras, and Tapia~Takaki]{Broz:2019kpl}
Broz, M.; Contreras, J.G.; Tapia~Takaki, J.D.
\newblock {A generator of forward neutrons for ultra-peripheral collisions:
${\textbf{n$\mathbf{_O^O}$n}}$}.
\newblock {\em Comput. Phys. Commun.} {\bf 2020}, {\em 253},~107181. [\href{http://dx.doi.org/10.1016/j.cpc.2020.107181}{CrossRef}]

\bibitem[Kryshen et~al.(2023)Kryshen, Strikman, and Zhalov]{Kryshen:2023bxy}
Kryshen, E.; Strikman, M.; Zhalov, M.
\newblock {Photoproduction of {$J/\psi$} with neutron tagging in
ultraperipheral collisions of nuclei at RHIC and at the LHC}.
\newblock {\em Phys. Rev. C} {\bf 2023}, {\em 108},~024904. [\href{http://dx.doi.org/10.1103/PhysRevC.108.024904}{CrossRef}]
\clearpage
\bibitem[Acharya et~al.(2023)]{ALICE:2023jgu}
Acharya, S. et~al. {[ALICE Collaboration]}
\newblock {Energy dependence of coherent photonuclear production of J/$\psi$
mesons in ultra-peripheral Pb--Pb collisions at $\sqrt{{\rm s}_{\mathrm{NN}}}=5.02$ TeV}.
\emph{J. High Energ. Phys.} {\bf 2023}, \emph{2023}, {119}. [\href{http://dx.doi.org/10.1007/JHEP10(2023)119}{CrossRef}]

\bibitem[Abdulhamid et~al.(2023)]{STAR:2023gpk}
Abdulhamid, M.{I.} et~al. {[STAR Collaboration]}
\newblock Exclusive $J/\psi$, $\psi (2s)$, and $e^+e^-$ pair production in ${\rm Au}+{\rm Au}$
ultraperipheral collisions at BNL Relativistic Heavy Ion Collider. \emph{Phys. Rev. C} {\bf 2024}, \emph{110}, {014911}. [\href{http://dx.doi.org/10.1103/PhysRevC.110.014911}{CrossRef}]


\end{thebibliography}
\end{document}